\documentclass[11pt,letterpaper]{article}
\usepackage{caption}
\usepackage{subcaption}
\usepackage{multicol}
\usepackage{graphicx}
\usepackage{amsmath}
\usepackage{amssymb}
\usepackage{svg}
\usepackage{float}
\usepackage[utf8]{inputenc}
\usepackage{mathtools}
\usepackage{setspace}
\usepackage[square,sort,comma,numbers]{natbib}
\usepackage{authblk}
\usepackage{hyperref}

\DeclarePairedDelimiter\floor{\lfloor}{\rfloor}

\title{Sampling Lattices in Semi-Grand Canonical Ensemble with Autoregressive Machine Learning}
\author[1]{James Damewood}
\author[1]{Daniel Schwalbe-Koda}
\author[1]{Rafael G\'omez-Bombarelli\thanks{rafagb@mit.edu}}
\affil[1]{
 Department of Materials Science and Engineering\\
 Massachusets Institute of Technology\\
 77 Massachusetts Avenue, Cambridge, MA 02319
}

\date{July 14, 2021}

\begin{document}
\citestyle{nature}
\bibliographystyle{naturemag}
\onehalfspacing

\maketitle

\section*{Abstract}

Calculating  thermodynamic potentials and observables efficiently and accurately is key for the application of statistical mechanics simulations to materials science. However, naive Monte Carlo approaches, on which such calculations are often dependent, struggle to scale to complex materials in many state-of-the-art disciplines such as the design of high entropy alloys or multicomponent catalysts. To address this issue, we adapt sampling tools built upon machine-learning based generative modeling to the materials space by transforming them into the semi-grand canonical ensemble. Furthermore, we show that the resulting models are transferable across wide-ranges of thermodynamic conditions and can be implemented with any internal energy model $U$, allowing integration into many existing materials workflows. We demonstrate the applicability of this approach to the simulation of benchmark systems (AgPd, CuAu) that exhibit diverse thermodynamic behavior in their phase diagrams. Finally, we discuss remaining challenges in model development and promising research directions for future improvements. 

\section{Introduction}

Reliable methods for the assessment of thermodynamic stability can accelerate materials design in at least two ways, one considering only energy and the other considering free energy. Identifying low-energy structures that are stable with respect to phase decomposition is needed to ensure that computer-designed materials are synthesizable and stable in operation conditions. In addition, including the role of temperature and entropy is required to understand phase transitions and to predict phase diagrams \textit{de novo.}

The difficulty in quantifying the free energy difference between phases arises because, in principle, the evaluation of potentials that govern phase stability requires a summation over all possible states of the system that satisfy the corresponding thermodynamic constraints. In practice, Monte Carlo (MC) methods can approximate equilibrium properties by identifying a relatively small number of representative system configurations from which ensemble averages can be estimated, and thus compute relative free energies and determine stable phases. The broad applicability of MC approaches has led to the development of numerous software packages specifically geared towards the materials domain \citep{CASM2021,Angqvist2019,Chang2019,Lerch2009,VandeWalle2002,VandeWalle2002a}. Generally, the most commonly implemented strategy to quantify phase stability is to: (1) consider a coarse-grained representation of a phase consisting of a supercell of fixed size and space group where states can be defined by a set of occupation variables ${\vec{S}}$ describing the atom at each site, (2) use a set of DFT (structure, energy) pairs to fit an empirical model that predicts the internal energy $U(\vec{S})$ of a state with occupancy  ${\vec{S}}$, and (3) draw samples from the equilibrium distribution defined by $U(\vec{S})$ using a Markov Chain. Each step of the chain, and thus the resulting representative configurations, is obtained through the stochastic proposal of a new state followed by an acceptance/rejection criteria determined by the relative probabilities of the new and previous states according to the equilibrium distribution.

While this method has demonstrated widespread utility, Markov Chain Monte Carlo (MCMC) \citep{Metropolis1953} requires serial computation, can suffer from critical slowing down near phase transitions, and results from simulations run at one set of fixed constraints are not generally transferable to other conditions. These issues can be partially mitigated by the design of specialized proposal/acceptance moves \citep{Swendsen1987,Wolff1989}, exchange between parallel simulations \citep{Swendsen1986}, and random-walks through the density of states \citep{Wang2001}, but many studies characterizing the mixing thermodynamics of complex, multi-component alloys often demand significant computational cost for large system sizes \citep{Widom2018}. 

These limitations have prompted the development of a number of novel MC methods specifically designed for multi-phase equilibria. Multi-cell Monte Carlo ($MC^2$) implements carefully designed proposal/acceptance steps such that atoms can be exchanged between separate supercells. The impact of phase interfaces on these finite-size simulations is significantly reduced as multiple phases can coexist across different cells \citep{Antillon2020,Niu2017,Niu2019}. Variance-constrained semi-grand canonical simulations rely on a new thermodynamic ensemble that can be leveraged to compute the free energies of systems within two-phase regions and improve the accuracy of recovered phase boundaries \citep{Sadigh2012}. Furthermore, Wang-Landau methods \citep{Wang2001} have been adapted to the materials domain and applied to characterize benchmark systems \citep{Takeuchi2017}.

Alternatively, machine learning approaches can be used to produce realistic high-likelihood samples from complex distributions without explicit parametrization, in so-called generative models \citep{Schwalbe-Koda2020,Gomez-Bombarelli2018}. The application of generative models to scientific calculations is a promising avenue to overcome the challenges of naive MC methods. Intuitively, these models are trained to draw samples by learning the typical values of the system's physical variables at equilibrium. A perfectly tuned model could then simulate the system by simply averaging over a batch of ML-proposed samples. Critically, when restricted to a class of exact-density models, this generative framework benefits from both a loss function relying on a variational estimate of the thermodynamic potential as well as reweighting \citep{Noe2019} and importance sampling techniques \citep{Nicoli2020} that can correct for sample distributions that deviate slightly from those at equilibrium.

The rigorous basis of these models and the explicit connection between exact likelihood and free energy have inspired a large number of physic-based applications. For continuous systems, exact-density flow models have been applied in reducing autocorrelations in lattice field theory \citep{Albergo2019,Kanwar2020,Pawlowski2020}, sampling free energy barriers of biomolecules \citep{Noe2019}, and studying relaxations of Ising models \citep{Li2018, Zhang2018}. In discrete cases, autoregressive models have been used to extract thermodynamics quantities \citep{Wu2019,Nicoli2020} and determine ground states \citep{Mcnaughton2020,Hibat-Allah2021} of spin models.

In this work, we introduce SEGAL (Semi-grand Ensemble Generation by Autoregressive Lattices), a generative approach to lattice simulations of phase stability in materials science. In particular, we demonstrate the applicability of exact-density generative models to the semi-grand canonical thermodynamic ensemble; assess model performance on well-known benchmark systems such as spin models, copper-gold and silver-palladium alloys; and extract estimates of phase stability of multi-component systems.

\section{Results}

\subsection{Autoregressive Sampling for Materials Simulation}
 
We seek to build a generative model that can successfully identify the representative states of the semi-grand canonical ensemble and their dependence on thermodynamic constraints, providing an alternative to traditional Monte Carlo approaches. We refer to this model as SEGAL (Semi-grand Ensemble Generation by Autoregressive Lattices). 

SEGAL associates each microstate the system can occupy with a predicted probability $P_{AR}$. Due to the discrete structure of the coarse-grained crystal representation, we decompose the probability of a particular decoration of the crystal prototype as a product of site probabilities that can represent any possible distribution over microstates. This mathematical decomposition requires defining an ordering over sites whereby the atomic identity of a particular site is dependent on its predecessors \citep{Wu2019, Nicoli2020}. Inspired by previous generative models that change the sampled distribution with temperature \citep{Singh2020,Noe2019,Dibak2020}, the dependencies between sites are also functions of the thermodynamic constraints, allowing the conditions to control the microstate probabilities,

\begin{equation}
    P_{AR}(\vec S | \Delta \mu, T) = \prod_i P(S_{i} | S_{j<i},\Delta \mu, T).
\end{equation}

\begin{figure}[htb]
\centering
\includegraphics[width=.75\linewidth]{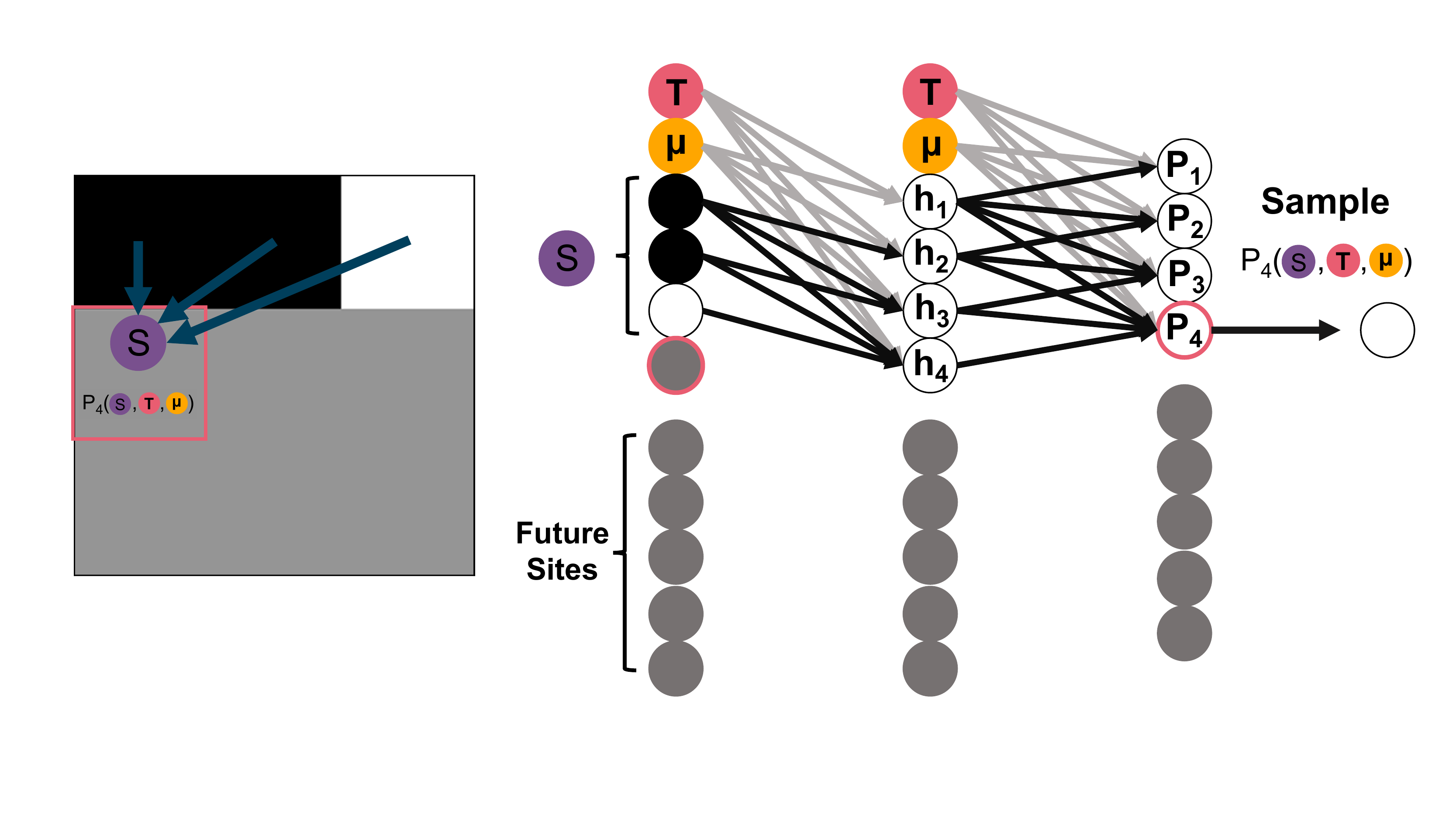}
\caption{Two-layer SEGAL architecture restricting dependence of all sites to previous sites and thermodynamic constraints. Each node shown in the figure represents a group of n neurons for an n-component system.}
\label{fig:Architecture}
\end{figure}

We parameterize these conditional probabilities using a neural network, whose general architecture is shown in Fig. \ref{fig:Architecture} and whose specific details per application are given in Figs. \ref{ising} to \ref{silverpalladium}. Therefore, the parameters of the network are trained to capture the underlying correlations of the atomic orderings. In order to generalize easily to arbitrary numbers of components and increase the capacity of the model, we represent each site $S_{i}$ as a vector with length equal to the number of components. New decorations of the lattice prototype can be drawn from the model by sequentially sampling each site $i$ from the categorical distribution $P(S_{i} | S_{j<i},\Delta \mu, T)$ such that, after sampling, $S_{i}$ is a one-hot encoded vector corresponding to the identity of the probabilistically chosen atom. The full state describing atomic labels over all sites is simply the concatenation of the $S_{i}$ vectors. Note that the first chosen site still has a dependence on the set $\Delta \mu$ and $T$.

\subsection{Ising Model in a Magnetic Field}

To demonstrate the use of SEGAL for a binary alloy, we first studied 10x10 periodic Ising spins in a magnetic field $B$. Through analysis of this model system, equivalences are drawn from spin variables to atomic site labels and from the magnetic field to the chemical potential difference. In particular, the long-range ordering of spins below the critical temperature is analogous to the opening of a two-phase miscibility gap in an alloy with unfavorable mixing. The internal energy function $U(\vec S)$ is the well-known nearest neighbor model with J=-1 in units of $k_b$:

\begin{equation}
    U(\vec S) = -\sum_{(i,j)\in \mathrm{NNs}} S_{i}\cdot S{j}
\end{equation}

\begin{figure}[htb]
\centering
\includegraphics[width=.9\linewidth]{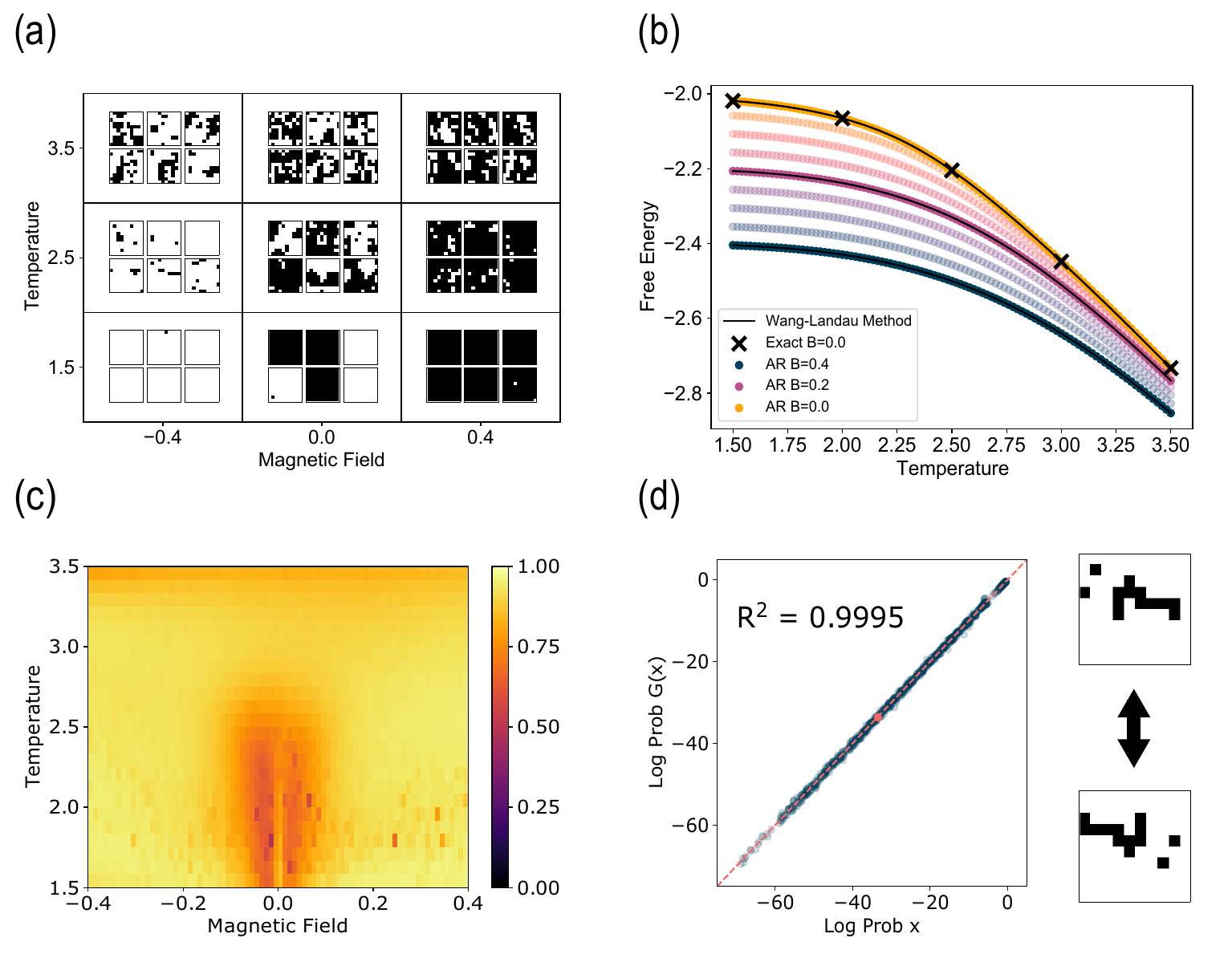}
\caption{SEGAL model for Ising alloy. (a) Samples from SEGAL under varying constraints $T \in [1.5,2.5,3.5] $, $B \in [-0.4,0.0,0.4] $. (b) Numerical comparison between self-normalized importance sampling (SNIS) from SEGAL, Wang-Landau method \citep{Belardinelli2007}, and exact solutions \citep{Beale1996,Pathria2011} with $B=0.0$. Colors with lower opacity are sampled at intermediate values of $B \in [0.05,0.10,0.15,0.25,0.30,0.35]$. SEGAL can interpolate over the whole training region. (c) Normalized effective sample size (NESS) over the training region. Estimates of NESS are taken with 10,000 samples each. (d) (left) Symmetry invariance of SEGAL model under an operation $G$. (right) An example of a transformed configuration with probabilities corresponding to the red dot on the left.}
\label{fig:IsingPanel}
\end{figure}

In the presence of a field $B$, an additional magnetic potential $\sum_{i} B \cdot S_{i}$ plays the role of chemical work $\sum_{i} \Delta \mu \cdot N_{i}$ for our model system. SEGAL is trained with $T \in [1.5,3.5]$ and  $B$ set to values $[-0.4,-0.2,0.0,0.2,0.4]$, a range over which both first-order and second-order phase transitions are known to occur. Qualitatively, samples from the trained network  exhibit behavior consistent with expectations (Fig. \ref{fig:IsingPanel}). At low temperature, ferromagnetic states are observed and demonstrate a first-order discontinuity at the critical magnetic field $B=0$. In addition, with increasing temperature, the samples demonstrate an order-disorder transition. Some magnetization values are never sampled, which is indicative of thermodynamically unstable alloy compositions that decompose into a linear combination of two more pure phases.

To quantitatively assess the validity of the model, we compared free energies estimated using self-normalized importance sampling on the output of SEGAL to those obtained from a Wang-Landau method that can interpolate between different temperatures but only at a fixed magnetic field \citep{Belardinelli2007}. When available, we also compared with exact results on finite size Ising models \citep{Kaufman1949,Beale1996,Pathria2011}. The reported Ising free energies are the mean of 10 independent calculations of $F(T,B)$ using 2000 samples each. Over the analyzed conditions, the differences in the free energy per site between the two methods are $O(10^{-4})$ and comparable in magnitude with the standard deviations of the 10 independent calculations of $F(T,B)$ (see Fig. \ref{fig:IsingErrors}). The total cost to train and sample this SEGAL model is $3\times10^{7}$ energy evaluations. When comparing to the exact values at $B=0$, the magnitude of the errors of SEGAL estimates are similar to the errors of the benchmark Wang-Landau algorithm \citep{Belardinelli2007} when ran for $10^{9}$ evaluations and restricted to zero magnetic field strength (see Fig \ref{fig:IsingTiming}). While this suggests that this SEGAL model is sample-efficient in learning the typical ensemble configurations, we note that this reduction in energy evaluations does not translate exactly to acceleration in wall clock time, because of the overhead of the neural network operations, the ability of the SEGAL to leverage batches to evaluate energies in parallel, and the Wang-Landau algorithm's exploitation of the local structure of $U$ to efficiently compute changes in energy between simulation steps. Though state-of-the-art exact density approaches have achieved accuracies of $\approx10^{-5}$ on 16x16 lattices at a single temperature \citep{Nicoli2020}, sacrificing optimal performance for generalizability over the space of constraints may have more practical utility in regimes when many sets of conditions are of interest, as is the case in predicting materials phase diagrams.

In order to provide another estimate on the quality of the self-normalized importance sampling, we measured the normalized effective sample size (NESS) over the conditions the model saw during training. While the effective sample size cannot be used to guarantee accurate model performance, it indicates where the model performs poorly. Over a wide range of conditions, SEGAL performs adequately, with a minimum NESS of 0.47. Areas with lower NESS give some intuition on the limitations of conditional generation. For instance, there are regions of lower NESS near the boundary of the training region, which is likely an artifact of the strategy used to sample different conditions during training. NESS is also lower near the first order phase transition where the ``typical'' configurations sampled by SEGAL change rapidly. Interestingly, above the critical temperature, performance no longer degrades significantly near $B=0$, which can be interpreted through the disappearance of the first-order phase transition.

The effective sample size is not a foolproof metric for performance, because a model suffering from mode-collapse --- that is, repeatedly producing only a very small set of unique outputs --- can still have high NESS. To address this concern, we further investigated potential mode collapse of the generative model. In particular, symmetry-related microstates must have the same unnormalized probability in the semi-grand canonical ensemble and that invariance should be preserved by SEGAL,

\begin{equation}
    P_{SG}(\vec S) = P_{SG}(G * \vec S),
\end{equation}

\noindent where U is invariant upon the operation G. A poorly regularized model could prefer samples with a particular translational or rotational orientation that would break the physical symmetry. In order to test our model, we generated samples over the full range of conditions and recorded their probabilities $P_{AR}(\vec S)$. We then applied a random symmetry operation and recorded the model probability of symmetry-adapted sample $P_{AR}(G*\vec S)$. If the generating field was non-zero, $G$ was composed of a random $C_{4}$ rotation composed with random translations in horizontal and vertical directions. If the B-field was 0.0 (10\% of the tests), an additional spin-flip operations was applied half the time. The $\log(P_{AR}(\vec S))$ and $\log(P_{AR}(G*\vec S))$ showed significant agreement ($R^{2} > 0.999$), suggesting that the model captures the underlying physical symmetries without the use of data augmentation or invariances being explicitly encoded in the network. One possible explanation for this performance is that the goal to accurately capture the ensemble under a range of ($B$, $T$) constraints forces the neural network towards varying regions of the systems order parameters including composition or site correlations. In this way, the training procedure may act as a natural regularizer of the generative model that incentivizes exploration and avoids mode collapse. Lastly, in Fig. \ref{fig:IsingAutoDiff} we explore how automatic differentiation\citep{Wang2020b} can be used to extract thermodynamic quantities by taking derivatives from the neural network predicted probabilities $P(\vec{S})$ instead of relying on fluctuations. 

\subsection{Ground States of CuAu}

\begin{figure}[!htb]
\centering
\includegraphics[width=.9\linewidth]{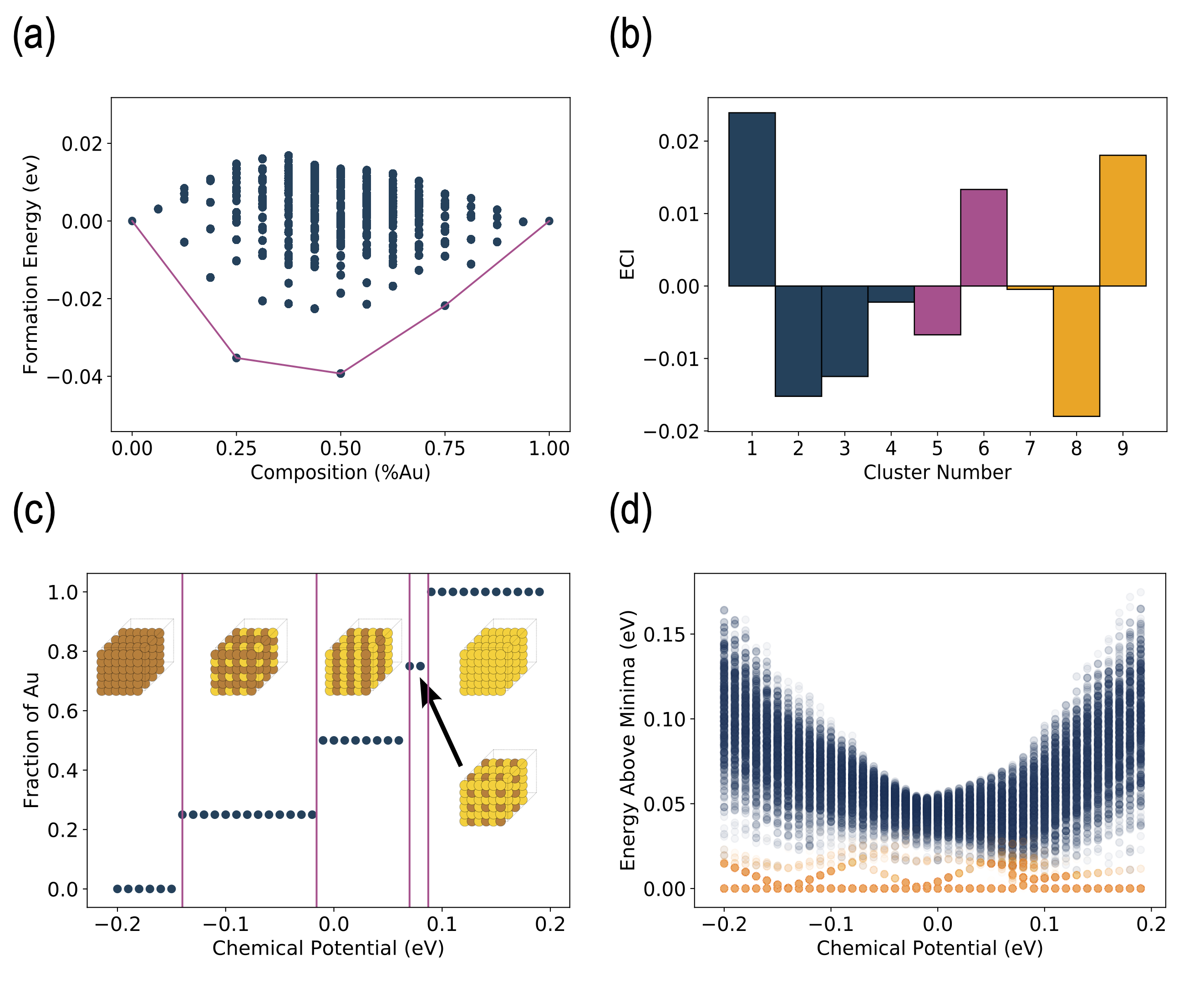}
\caption{(a) Formation energies of all $2^{16}=65,536$ possible decorations of 16-site fcc prototype lattice according to the model cluster expansion. (b) Effective cluster interactions (ECI) of two-body (blue), three-body (purple), and four-body (orange) clusters obtained using CLEASE \citep{Chang2019}. (c) Ground states sampled as a function of composition. Cells are expanded by nine times to aid visualization. Purple lines denote expected transitions based on the cluster expansion energy function. (d) Comparison of sampling of grand potential minima with SEGAL (orange) and random search (blue).}
\label{fig:CuAuPanel}
\end{figure}

In order to test the ability of SEGAL to detect low internal energy phases on realistic materials, we analyzed its performance detecting the stable ordered structures in a copper-gold alloy, a widely studied system for MC algorithms and software \citep{Fontaine1994,Lu1991,Ozolins1998,Zhang2014}. As is standard in materials science workflows, we trained a cluster expansion $U(\vec{S})$ model  to predict the energy of new decorations of fcc lattices with the aid of the CLEASE \citep{Chang2019} package. Density Functional Theory (DFT) energies were computed for a total of 68 training structures with fcc lattices of various sizes, generated using a combination of random search, probe structures, and simulated annealing. We observed that including all the data resulted in a degradation in the ability of the cluster expansion to accurately fit the low energy structures relevant for the ground state search. Previous work also found that depending on the application context, cluster expansion performance can be sensitive to the choice of training data \citep{Kleiven2021}. The highest prediction accuracy for low energy structures was obtained using a set of 40 training examples with formation energies below 0.02 eV/atom, resulting in a final cluster expansion with a leave one out cross-validation (LOOCV) score of 8.6 meV/atom. The effective cluster interactions (ECI) parameters and convex hull for a 16-site supercell are shown in Fig. \ref{fig:CuAuPanel}, predicting $\mathrm{Cu_{3}Au}$, $\mathrm{CuAu}$, and $\mathrm{Au_{3}Cu}$ as stable intermetallics. The training structures included the pure phases as well as the $\mathrm{CuAu}$ and $\mathrm{Cu_{3}Au}$ ground states, but not the stable $\mathrm{Au_{3}Cu}$ structure.

To sample ground states of varying composition, SEGAL is trained on a 16-site fcc lattice prototype over a range of chemical potential differences bounded by values where the pure phases are stable, $\Delta\mu \in [-0.24,0.24]$. The temperature was steadily decreased over each epoch in a simulated annealing-based approach to increase the likelihood of converging to the correct structures. A similar method was employed by Wu et al. to minimize the energy of spin systems \citep{Wu2019}. Note that in contrast to SEGAL, the minimization of energy alone would only result in the detection of the structure with minimum formation energy (CuAu). The total number of energy evaluations required to train SEGAL on the CuAu system is 1,000,000, which exceeds the number of possible states on the 16-site lattice (65,536), but the resulting model can still be used to examine SEGAL's behavior in the context of real material system.

Once trained, modifying the chemical potential difference allows SEGAL to sample stable alloy structures of varying composition, successfully identifying pure phases as well as the $\mathrm{Cu_{3}Au}$, $\mathrm{CuAu}$, and $\mathrm{CuAu_{3}}$ intermetallics. Futhermore, when stability is determined by the minimum value of the grand potential at 0 K over a batch of 1000 samples, the critical chemical potentials between stable structures closely match those predicted by the convex hull of the cluster expansion, suggesting that SEGAL has learned to approximate the location of phase transitions. 

We observe a greater degree of mode collapse than with the Ising model case, as the model finds $\mathrm{Cu_{3}Au}$, $\mathrm{CuAu}$, and $\mathrm{CuAu_{3}}$ ground states with degeneracy 2, 1, and 2, respectively, where the exact values determined through a brute force enumeration are 5, 7, and 5. The increased difficulty of this task could be due to the more complicated symmetry relationships between ground states or the convergence of training temperature to 0 K, which reduces the regularization effects of temperature variability, from which the Ising SEGAL model may have benefited more significantly.

We compared the effectiveness of the trained SEGAL model to a benchmark random algorithm that samples all configurations with equal frequency by recording the percent of samples that correctly identify the grand potential minima at 0 K. During the test, 1,000 samples were drawn from each method at 40 separate values of $\Delta\mu$. In total, only 1 of the 40,000 random samples identified the correct structure, whereas $74\%$ of the SEGAL samples correspond to the grand potential minima. Therefore, we conclude that SEGAL is capable of extracting stability-relevant thermodynamic information from a model of a real material's internal energy after being trained. Similar to the observation of the Ising model's NESS, the lowest probability of sampling the correct structure occurs as $\Delta\mu$ approaches phase transitions, where two competing ground states have very similar grand potentials and  SEGAL-generated structures must rapidly switch between phases. This effect is largest in the case of $\mathrm{Au_{3}Cu}$, which is only stable for a narrow range of chemical potentials (see Fig. \ref{fig:CuAu_GSProb}).

\subsection{AgPd Alloy}

We further explored the ability of SEGAL to capture the physics of a real metal alloy at finite temperature. As an example, we considered a 27-site fcc prototype (3x3x3 supercell)  of silver and palladium, whose phase diagram features a miscibility gap extending to temperatures of up to 600 K. Below the top of the miscibility gap, unfavorable mixing interactions cause ranges of alloy compositions to be thermodynamically unstable. The gap exhibits a characteristic asymmetry, as palladium is highly soluble in silver, but silver has virtually no solubility in palladium at low temperatures \citep{Ghosh1999,Dinsdale2008}. 

A cluster expansion approximation of the formation energy $U_{CE}$ was built using a dataset of 625 AgPd structures from the ICET \citep{Angqvist2019} tutorial database and obtained a 10-fold cross validation error of 2.2 meV/atom. SEGAL was trained using $U_{CE}$ over a temperature range of [200 K,900 K] extending within and above the expected miscibility gap. Benchmark semi-grand canonical Markov Chain Monte Carlo simulations using the same cluster expansion were run using CLEASE. In order to show the flexibility of SEGAL with regard to the energy model $U$, we also trained a crystal graph convolutional model for the formation energy $U_{CGC}$ over the same dataset, which achieved a test error of 1.34  meV/atom \citep{Xie2018}. For the crystal graph convolutional model, we wrote our own CGC MCMC implementation to obtain reference values.

Results from self-normalized importance sampling and the Markov Chain estimates show strong numerical agreement across multiple temperatures for both energy models, with deviations in composition on the order of $10^{-3}$ (see Fig. \ref{fig:AgPd Errors}). These errors are sufficiently small to recover the physical properties and phase stability of the alloy over the training region. At 250 K, the discontinuity in compositions indicates thermodynamically unstable compositions and confirms the presence of the two-phase region, separating a nearly pure Pd phase and a 60/40 mixture of Pd and Ag. At 750 K, both methods show continuous variation in composition with chemical potential, suggesting that the top of the miscibility gap has been exceeded. Importantly, SEGAL is applicable as a sampling method for both $U_{CGC}$ and $U_{CE}$ potentials, and can be readily generalized to any newly developed models for alloy energy.

The normalized effective sample size of SEGAL is reasonable over a large range of conditions, but indicates lower performance near the critical values of $\Delta\mu$, at which the discontinuity in composition is observed and the typical lattice configurations at equilibrium change rapidly. These uncertainties near phase transitions can introduce deviations in the bounds of the two-phase region such as those observed at 250 K for the $U_{CGC}$ model. We further note that above the miscibility gap ($\approx$600 K), stable compositions change more continuously, and the subsequent decrease in the NESS metric is significantly less pronounced. By identifying regions of constraint space where typical states of the system change rapidly, NESS calculations of SEGAL models show some promise at the automatic detection of phase transitions.

\begin{figure}[H]
\centering
\includegraphics[width=\linewidth]{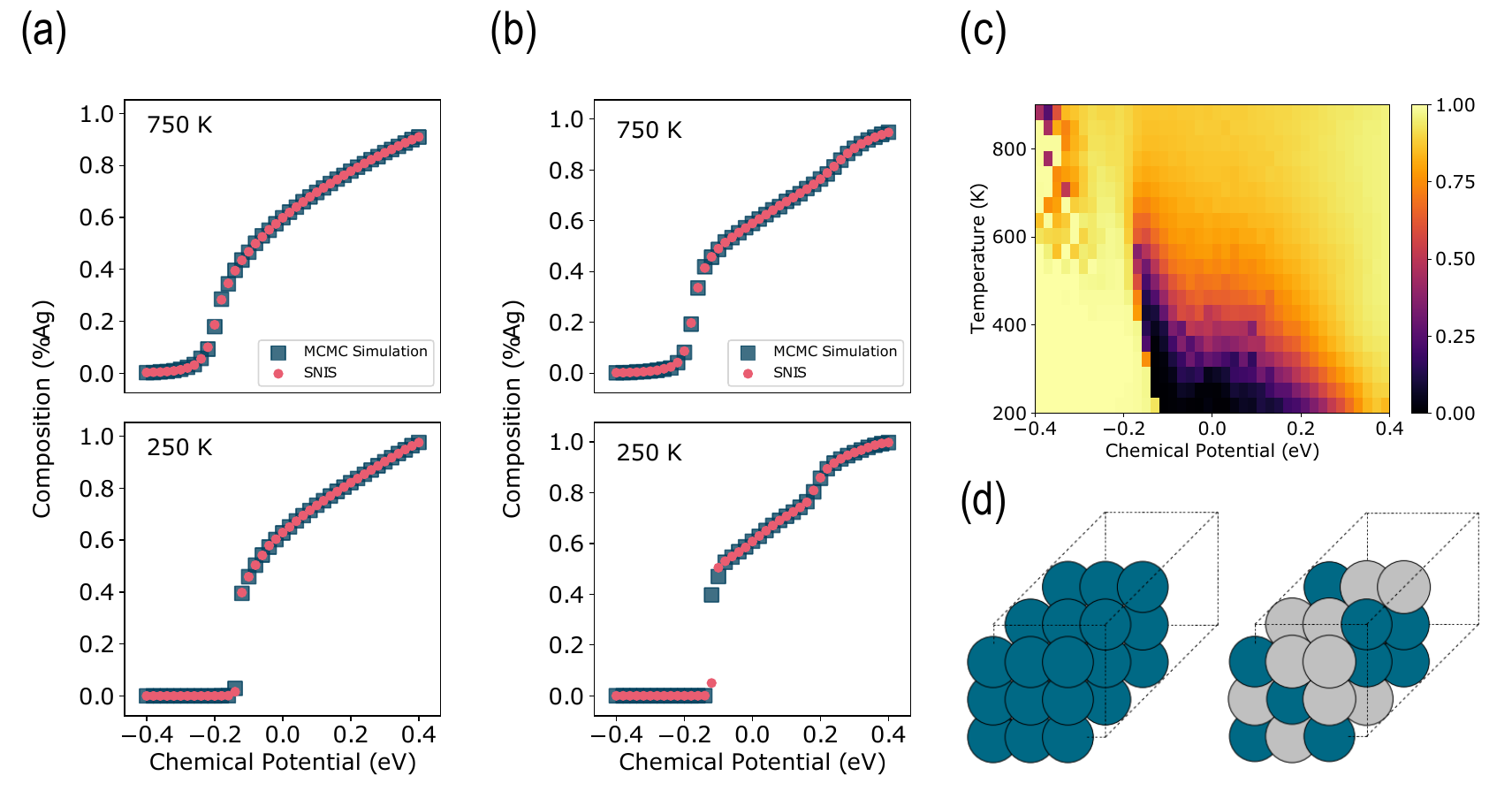}
\caption{SEGAL applied to 27-site AgPd alloy. (a),(b) Composition vs. $\Delta \mu$ computed with MCMC and self-normalized importance sampling (SNIS) at temperatures of 750 K (top) and 250 K (bottom) using (a) cluster expansion  and (b) crystal graph convolution models for $U$. (c) Normalized effective sample size of SEGAL over the training region. Estimates of NESS are taken with 10,000 samples each. (d) Samples from SEGAL at $T = 250 \mathrm{ K}, \Delta \mu = -0.16$ and $T = 250 \mathrm{ K}, \Delta \mu = -0.04$ (right).}
\label{fig:AgPdPanel}
\end{figure}

\subsection{Predicting Phase Stability}

\begin{figure}[H]
\centering
\includegraphics[width=\linewidth]{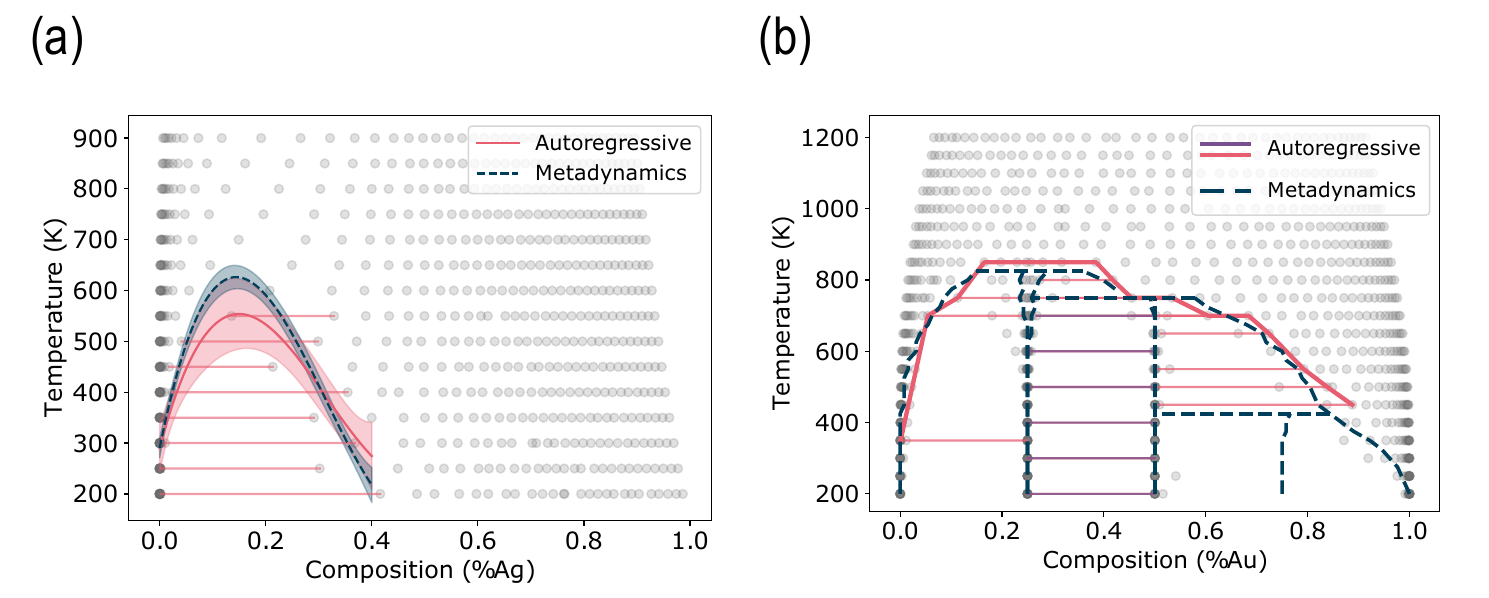}
\caption{Prediction of phase diagrams from SEGAL compared with metadynamics benchmark for (a) 125-site AgPd model and (b) 128-site AuCu model. For (a), error bars are computed from the uncertainty on the polynomial fit. For (b), purple lines show two-phase region for $\mathrm{AuCu{3}}$ and $\mathrm{AuCu}$. Pink lines show two-phase equilibria between ordered compounds and disordered solid solution. }
\label{fig:PhaseDiagrams}
\end{figure}

Finally, we give examples on how the SEGAL model can be used to extract information on phase stability. To reduce the artificial effects of a finite simulation cell, we trained SEGAL on larger cells for the AgPd (125-site) and CuAu (128-site) systems. After drawing 5,000 samples from the AgPd model for 15 temperatures between  200 and 900 K, and 41 values of $\Delta\mu$ between  -0.4 and +0.4 eV, a region of thermodynamically unstable compositions was visible and attributed to the miscibility gap. The top of the gap was estimated using an alpha shape algorithm \citep{Bellock2021}, a generalization of the convex hull (see Fig.  \ref{fig:AgPd_AlphaShape}). The boundary of the gap was computed using a polynomial fit to the points exhibiting the greatest discontinuity in composition at or below the critical temperature. For the CuAu model, 5,000 samples were drawn at 21 temperatures from 200 K to 1200 K and 41 values of $\Delta\mu$ from -0.24 eV to +0.24 eV. Observed discontinuities in stable compositions suggested the presence of a $\mathrm{Cu_{3}Au}-CuAu$ two-phase region for temperatures below 700 K. Estimated bounds were determined from the maximum difference in composition between $\Delta\mu$ values separated by 0.024 eV, restricted to the composition range $0.2 < \%Au < 0.6$. Based of previous work of Takeuchi et al. \citep{Takeuchi2017}, bounds for order-disordered two-phase regions were estimated by locating the temperature with maximal heat capacity $T_{C}$ for each constant value of $\Delta\mu$ and approximating the bounds of the two-phase regions as the compositions at $(T_{C},\Delta\mu-\delta)$ and $(T_{C},\Delta\mu+\delta)$ with $\delta = 0.012$ eV (See Fig. \ref{fig:CuAu_HeatCap}). Results for both systems agree favorably with reference metadynamics simulations (Fig. \ref{fig:PhaseDiagrams}).

The total number of energy evaluations using the cluster expansion model required to train and sample the AgPd and CuAu models with SEGAL were $10^7$ and $2.6\times10^{7}$ respectively. The baseline metadynamics simulations run over the same temperature range required $3.9\times10^{7}$ (AgPd) and $1.8\times10^{8}$(CuAu) energy evaluations. However, we note that due to the increased accuracy of the metadynamics simulations, highlighted by the detection of the $\mathrm{Au_{3}Cu}$ phase, these values are not directly comparable. The NESS values of these larger models (see Fig. \ref{fig:LargeNESS}) exhibit many of the similar trends as previous experiments such as low values in the vicinity of phase transitions. In contrast, NESS values in the disordered phases are $O(10^{-3}-10^{-2})$, significantly lower than those observed for the smaller alloys and Ising system. As a result, the efficient scaling of SEGAL models to large cell sizes of complex alloys is an outstanding challenge, but holds promise for the simulation of multi-component systems.

\section{Discussion}

We have shown that general-purpose generative models for statistical physics can be readily modified for applications computing thermodynamic quantities in materials science. In particular, transforming to the semi-grand canonical ensemble avoids interfaces between competing phases and allows for a greater control over the exploration of experimental order parameters such as composition and atomic ordering. Furthermore, a single model with no training examples from previous simulations can generalize across a wide range of constraints and accurately determine thermodynamic potentials, observables, and stable phases. SEGAL does not restrict the form of the potential $U(\vec{S})$ in any way and can be trained with Crystal Graph Convolution networks \citep{Xie2018} or other approaches capable of modeling complex multi-component systems \citep{Liu2021}.  As a result, generative models have the potential to become a useful tool alongside standard lattice simulation techniques. 

While the approach is promising, a number of algorithmic improvements could improve its scalability and performance. The current architecture is more sample efficient than baseline methods, but does not scale to cell sizes comparable to those of typical simulations, required to increase the precision of the final estimates and reduce finite size effects. Possible architecture changes could include implementing autoregressive network using graph convolutional layers to utilize the symmetry of the crystal system \citep{Wu2019} or exploiting the local structure of the energy model to improve the scalability of the generation process \citep{Pan2021,Dai2020}. Another crucial step is to refine SEGAL's sampling performance near phase transitions. SEGAL's ability to identify these regions through a change in ``typical'' states and the associated decrease in NESS values could allow for modified training strategies. In particular, training batches can be more frequently focused in regions with low NESS so that additional examples can help the model to improve in cases where the learning task is difficult. Alternatively, SEGAL could be supplemented with standard MCMC simulations run with constraints close to the critical values of temperature and chemical potential, or with strategies to account for exponentially suppressed configurations that increase the variance of importance sampling estimates \citep{Wu2021}.

\section{Methods}

\subsection{Thermodynamics Ensembles}

To draw samples from a particular equilibrium ensemble, lattice Monte Carlo simulations must be run under a chosen set of thermodynamic constraints. In the canonical ensemble, temperature and composition are fixed and system configurations are sampled according to their relative Boltzmann weight $\propto e^{-\frac{U}{k_{b}T}}$. Free energies obtained through this approach can characterize a wide range of phenomena in statistical physics. However, when investigating multi-component materials thermodynamics, the free energy minimum can be achieved by any linear combination of phases that satisfies the composition constraints. Therefore, at equilibrium, multiple phases can coexist in a manner that cannot be represented with a single fixed lattice prototype without introducing phase boundaries. The presence of these multi-phase regions must then be inferred from non-convex regions of the free energy as a function of composition that was observed in the simulation. In order to alleviate this challenge, materials scientists often work in the grand-canonical ensemble with fixed chemical potentials and temperature. In this ensemble, for each set of constraints only a single phase will be present at equilibrium, except at the critical values where phase transitions occur. As a result, simulations avoid multi-phase equilibria and are more well-suited to a single lattice cell. While generative adversarial networks have been applied to the grand canonical ensemble in the context of scalar field theory \citep{Zhou2019}, most previous exact-density approaches \citep{Noe2019, Wu2019, Nicoli2020} have modeled the canonical ensemble. As such, the applicability of these new methods to the materials community is an open question.

The grand potential and resulting microstate probabilities can be derived for a system of $i$ species through a Legendre transform of the canonical ensemble. With a fixed total number of sites $\sum_{i}N_{i} = N_{tot}$, the system is in the semi-grand canonical ensemble and is determined by a set of $i-1$ chemical potential differences, $\Delta \mu_{i} = \mu_{i} - \mu_{0} $, and the temperature.

\begin{equation}
    \Phi(T,\{\Delta\mu_{i}\})  = F -\sum_{i \neq 0}\Delta\mu_{i} N_{i} = U - TS - \sum_{i \neq 0}\Delta \mu_{i} N_{i}
\end{equation}

\begin{equation}
    P_{SG}(\vec S | T,\{\Delta\mu_{i}\} ) = \frac{e^{[-U(\vec S) + \sum_{i \neq 0} \Delta \mu_{i} N_{i} (\vec S)]/k_{b}T}}{Z_{SG}}
\end{equation}

 The relative probabilities, and thus, the representative configurations the system occupy at equilibrium change in response to the above constraints. In particular, varying the chemical potential differences results in driving forces to introduce changes in composition, and increasing the temperature leads to a greater contribution to the grand potential from configurational entropy and greater system disorder. We demonstrate the dependence of composition on chemical potential for a toy system in Fig. \ref{fig:Legendre}).

\subsection{Training}

If the sampler was perfect, all microstates configurations would appear with the same relative probabilities as they do in the studied thermodynamic ensemble. One approach to encourage the model probability distribution to converge on the correct values is to minimize the KL divergence, a measure of the difference between two probability distributions, between the model and the ensemble $KL(P_{AR}|P_{SG})$. It can be shown that (see Fig. \ref{si_loss_function}) the resulting minimization objective can be expressed as:

\begin{equation}\label{Loss}
      KL(P_{AR}|P_{SG}) = \mathbf{E}_{AR} [\frac{U(\vec S)}{k_{b}T} + \log[P_{AR}(\vec S)]- \frac{\Delta\mu N (\vec S)}{k_{b}T} ] = \frac{\Phi_{AR}(T,\{\Delta\mu_{i}\})}{k_{b}T}
\end{equation}

The true grand potential is the minimum of $\langle U - ST - \sum_{i \neq 0} \Delta\mu_{i}N_{i}\rangle_{SG} $ for all possible probability distributions over microstates and will provide a lower bound on the training loss function such that $\Phi_{AR} \geq \Phi_{SG}$. While Eqn.\eqref{Loss} is not differentiable due to the discrete, stochastic sampling step, gradients can be estimated through \citep{Wu2019}:

\begin{equation}
    \nabla_{\phi} KL(P_{AR}|P_{SG}) = \mathbf{E}_{AR} [\log(\frac{P_{AR}(\vec S)}{\hat{P}_{SG}(\vec S)})\nabla_{\phi}\log(P_{AR}(\vec S))]
\end{equation}

\begin{equation}
    \log(\hat{P}_{SG}(\vec S)) = -\frac{U(\vec S)}{k_{b}T} + \frac{\Delta\mu N (\vec S)}{k_{b}T}+\frac{\hat{\Phi}_{AR}(T,\{\Delta\mu_{i}\})}{k_{b}T}
\end{equation}

\noindent where $\hat{\Phi}_{AR}$ is an estimate of Eqn.\eqref{Loss} over the whole batch of samples. Intuitively, the model will seek to lower the likelihood of configurations for which $P_{AR} > \hat{P}_{SG}$ and increase the likelihood of configurations for which $P_{AR} < \hat{P}_{SG}$. Because $U(\vec S)$ is not required to be differentiable, a wide range of standard energy models can be easily incorporated into this approach.

Training SEGAL does not require any example configurations, only an energy function $U(\vec S)$ to model. Batches of samples are iteratively drawn and used to estimate the loss function and update model parameters. As training continues, the estimated grand potential $\hat{\Phi}_{AR}$ decreases towards the true minimum $\Phi_{SG}$ and the relative probabilities of the samples approach their equilibrium values. We found multiple procedures could be implemented in order to effectively allow the model to capture the condition-dependent equilibrium distribution. The chemical potential differences ${\Delta\mu_{batch}}$ and temperature $T_{batch}$ of each batch could be set randomly using a uniform distribution within the bounds being investigated $T_{batch} \in [T_{min},T_{max}]$, $\Delta \mu_{batch} \in [\mu_{min},\mu_{max}]$ or set to specific values chosen as hyperparameters. Training can be stabilized by computing the loss over several sets of conditions $[T_{batch},\Delta \mu_{batch}]$ simultaneously before updating parameters. In this case, estimates of $\hat{\Phi}_{AR}$ are computed separately over constant conditions. In addition, because the magnitude of thermodynamic potentials can differ significantly depending on the constraints, when combining samples generated under different conditions the gradients were further normalized by the absolute value of $\frac{\hat{\Phi}_{AR}(T,\{\Delta\mu_{i}\})}{k_{b}T}$. Following the learning procedure, the model can draw samples over the entire range of conditions it was exposed to during training.

 \subsection{Neural Importance Sampling}
 
 Despite the physics-informed training procedure, generative models will not achieve perfect performance for any ensemble and estimates of thermodynamic observables can be significantly biased \citep{Wu2019,Nicoli2020}. However, if the probability of the proposed samples $P_{AR}$ is known exactly, the statistical power of numerical estimates can be improved by weighting samples using the relation:

\begin{equation}
    \mathbf{E}_{SG}[O(\vec S)] = \mathbf{E}_{AR}[\frac{P_{SG}(\vec S)}{P_{AR}(\vec S)}O(\vec S)]
\end{equation}
  
where, for example, samples that appear more frequently in the generated distribution than in the target distribution are given less weight to compensate for their increased rate of appearance. While the normalizing constant of $P_{SG}$ is unknown in many practical problems, samples can be still be treated as a well-designed proposal distribution for a Markov Chain \citep{Albergo2019} or used as a biasing distribution for histogram reweighting \citep{Noe2019}. Nicoli et al. \citep{Nicoli2020} introduced the use of generative models with Self-Normalized Importance Sampling (SNIS), which offers the added benefit of providing estimates of both normalizing constants and observables. Defining $w(\vec S)$ as the unnormalized ensemble probability divided by the generative model probability $P_{AR}$:

\begin{equation}\label{SNIS}
    \mathbf{E}_{SG}[O(\vec S)] = \mathbf{E}_{AR}[\frac{w(\vec S)}{Z_{SG}}O(\vec S)]
\end{equation}

\begin{equation}
    Z_{SG} = \mathbf{E}_{AR}[w(\vec S)] = \mathbf{E}_{AR}[e^{[-U(\vec S) +  \Delta \mu N (\vec S)]/k_{b}T}/P_{AR}(\vec S)]
\end{equation}
 
Because an estimate of $Z_{SG}$ must be used in Eqn.\eqref{SNIS}, SNIS is still biased in practice, but the biases can be substantially smaller than those achieved by simply averaging over samples of the generative model. One metric to evaluate this approach is the effective sample size (ESS), which provides an estimate on the number of samples from the true target distribution required to match the performance of the SNIS. The ESS can be normalized (NESS) to evaluate the typical quality of generated samples when compared with the target distribution.

\begin{equation}
    NESS = \frac{1}{n} \frac{\sum_{i}^{n}{w_{i}^{2}}}{(\sum_{i}^{n}{w_{i}})^{2}}
\end{equation}

Note that if the generated distribution closely resembles the target distribution and all $w_{i}$ are close to $Z_{SG}$, the NESS will approach 1. As the generated distribution deviates from the target and the variation in $w_{i}$ increases, the NESS will approach 0. 

\subsection{Density Functional Theory calculations}

DFT calculations were carried out using the Vienna Ab-initio Simulation Package (VASP), \cite{Kresse1996,Kresse1996a} v. 5.4.4, within the projector-augmented wave (PAW) method.\cite{Blochl1994,Kresse1999} The Perdew–Burke–Ernzerhof (PBE) functional within the generalized gradient approximation (GGA) \cite{Perdew1996} was employed as the exchange-correlation functional, including dispersion corrections through Grimme's D3 method. \cite{Grimme2010,Grimme2011} The kinetic energy cutoff for plane waves was restricted to 520 eV. Integrations over the Brillouin zone were performed using Monkhorst-Pack $k$-point meshes\cite{Monkhorst1976} with a uniform density of 64 $k$-points/\AA$^{-3}$. A stopping criterion of $10^{-6}$ eV was adopted for the electronic convergence within the self-consistent field cycle. Optimization of unit cell parameters and atomic positions was performed until the Hellmann–Feynman forces on atoms were smaller than 10 meV/\AA.

\subsection{Data and Code availability}
The algorithms reported in this work, trained models, and the DFT training data used to fit the $U(\vec S)$ models, are available under at https://github.com/learningmatter-mit/Segal.  

\subsection{Acknowledgements}
This work was supported by ARPAe DIFFERENTIATE (Award No DE-AR0001220) and by Zapata Computing Inc. JKD acknowledges support from the National Defense Science and Engineering Graduate Fellowship. D.S.-K. was additionally supported by the MIT Energy Fellowship.

\bibliography{SEGAL_June.bib}

\newpage

\renewcommand\thefigure{S\arabic{figure}}    
\renewcommand*{\thesubsection}{S\arabic{subsection}}
\setcounter{figure}{0}    
\setcounter{subsection}{0}

\section*{Supplementary Information}

\subsection{Derivation of Loss Function}
\label{si_loss_function}
The $KL$ divergence between the model and the true distributions can be expressed as follows:

\begin{equation}
    KL(P_{AR}||P_{SG}) = \mathbf{E}_{AR}[\log(\frac{P_{AR}}{P_{SG}})]
\end{equation}

\begin{equation}
    \mathbf{E}_{AR}[\log(\frac{P_{AR}}{P_{SG}})] = \mathbf{E}_{AR}[\log(P_{AR}) - \log(P_{SG}) ]
\end{equation}

\begin{equation}
    = \mathbf{E}_{AR} [\log[P_{AR}(\vec S)]+\frac{U(\vec S)}{k_{b}T} - \frac{\Delta\mu N (\vec S)}{k_{b}T} + \log(Z_{SG})]
\end{equation}

where $\log(Z_{SG})$ is a constant that can that be removed without affecting the optimization.

\begin{figure}[H]
    \centering
    
    \begin{subfigure}[t]{0.45\textwidth}
        \centering
        \includegraphics[width=\linewidth]{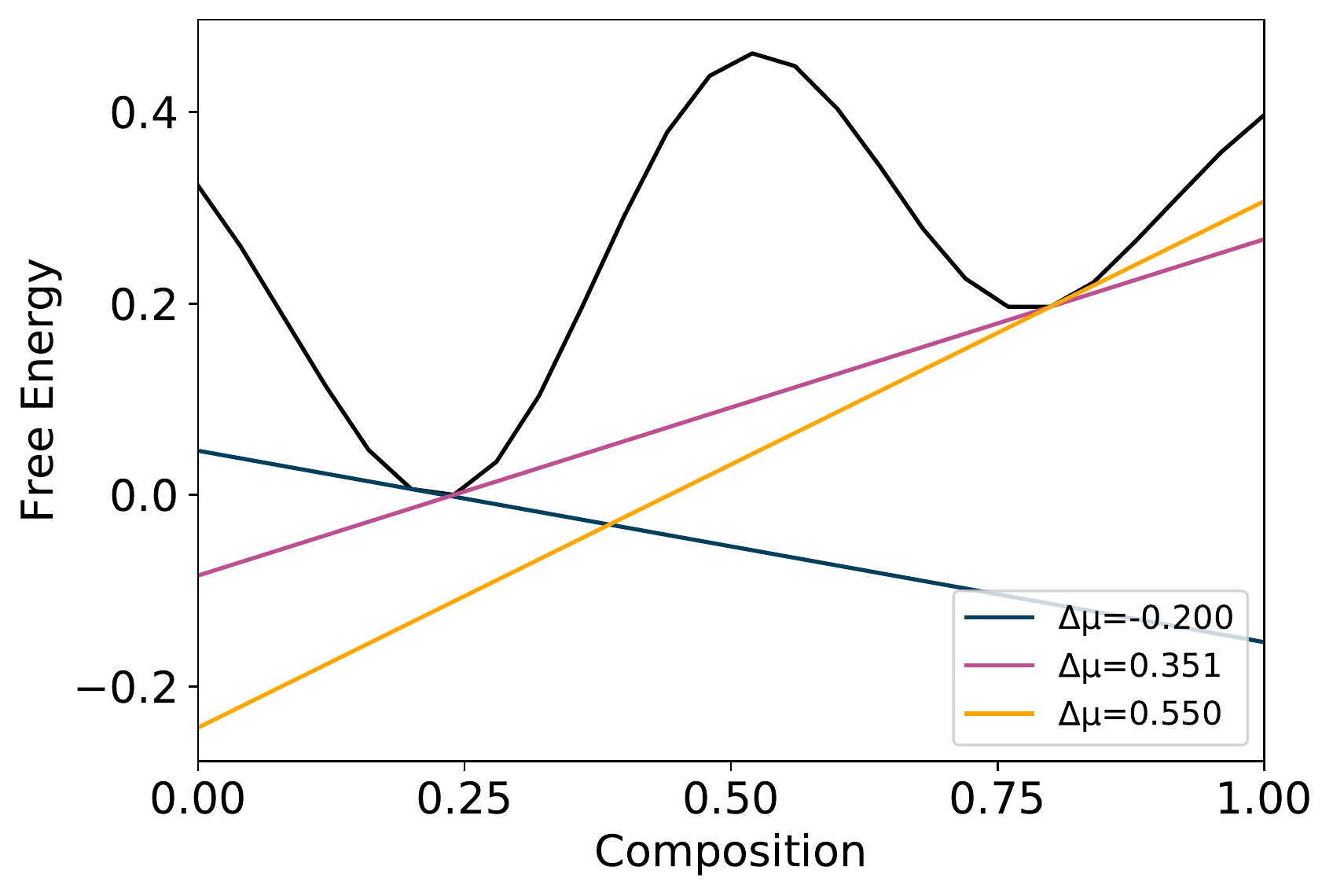} 
        \caption{} 
    \end{subfigure}
    \begin{subfigure}[t]{0.205\textwidth}
        \centering
        \includegraphics[width=\linewidth]{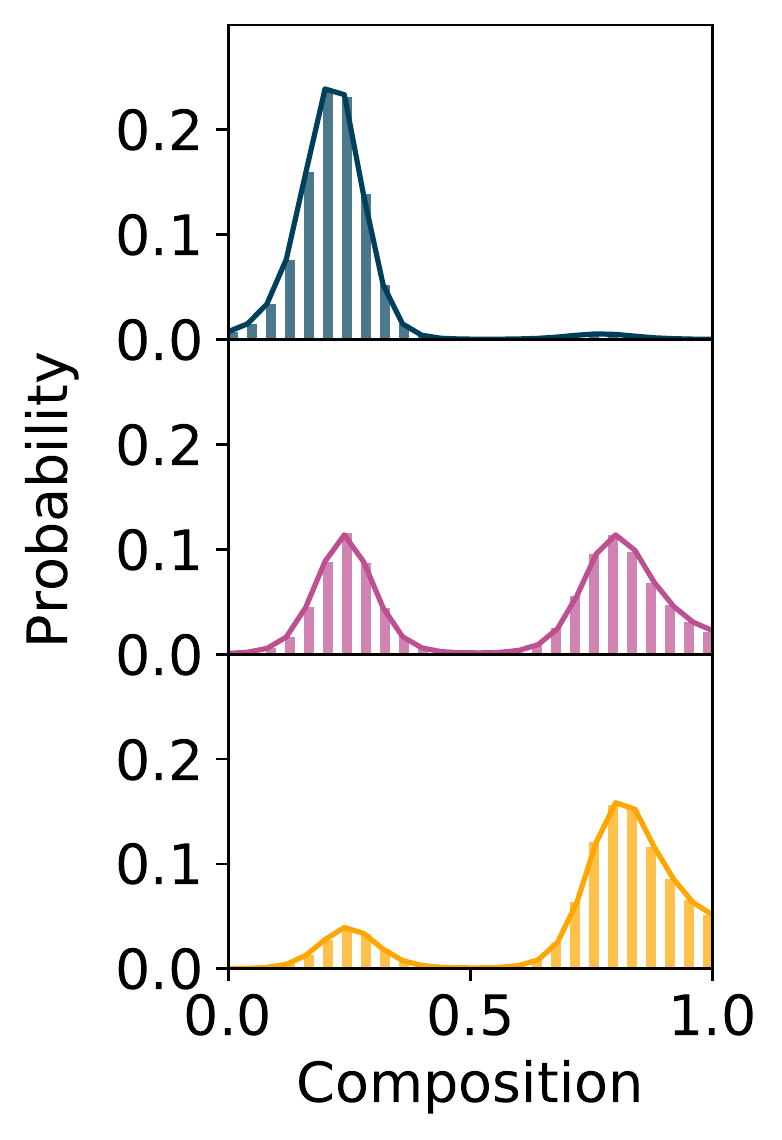} 
        \caption{} 
    \end{subfigure}
    
    \caption{Semi-grand canonical thermodynamics of two-component system with N = 25. (a) Free energy per site with varying composition computed using enumeration of all states and tangent construction of Legendre transform at varying values of $\Delta \mu$. (b) Composition probability histograms for values of $\Delta \mu$ of -0.200 (top), 0.351 (middle), and 0.550 (bottom) obtained through enumeration of all states (lines) and self-normalized importance sampling (SNIS) (bars).}
    
    \label{fig:Legendre}
\end{figure}

\subsection{Architecture Details}
\label{architecture}
In this section, we describe the structure of the neural network architectures used in this work. We represent each site variable $S_{i}$ as a categorical one-hot over the number of possible components in the system. $\vec{S}$ is then the concatenation of $S_{i}$. Therefore, for a 50-site binary system, $\vec{S}$ is 100 dimensional.

The simplest autoregressive layer is a weight matrix $W$ with masked parameters such that the imposed dependence between the variables is preserved. With $T$ and $\Delta \mu$ constraints appended to the start of $\vec{S}$, the non-zero components of a single layer $W_{ij}$ include $j \leq 2*\floor{\frac{i}{2}}$. The weight matrix of the second layer can be expanded with non-zero terms $W_{ij}$, $j \leq 2*(\floor{\frac{i}{2}}+1)$. Element-wise activation functions can be included after the application of weight matrices.

Additional layers to the autoregressive networks can be added with the same form as the second layer. The depth of all models is specified in the corresponding experimental details section.

\subsection{Ising Alloy Experimental Details}
\label{ising}

\begin{figure}[H]
    \centering
    \begin{multicols}{2}

    \hfill
    \begin{subfigure}[t]{0.45\textwidth}
        \centering
        \includegraphics[width=\linewidth]{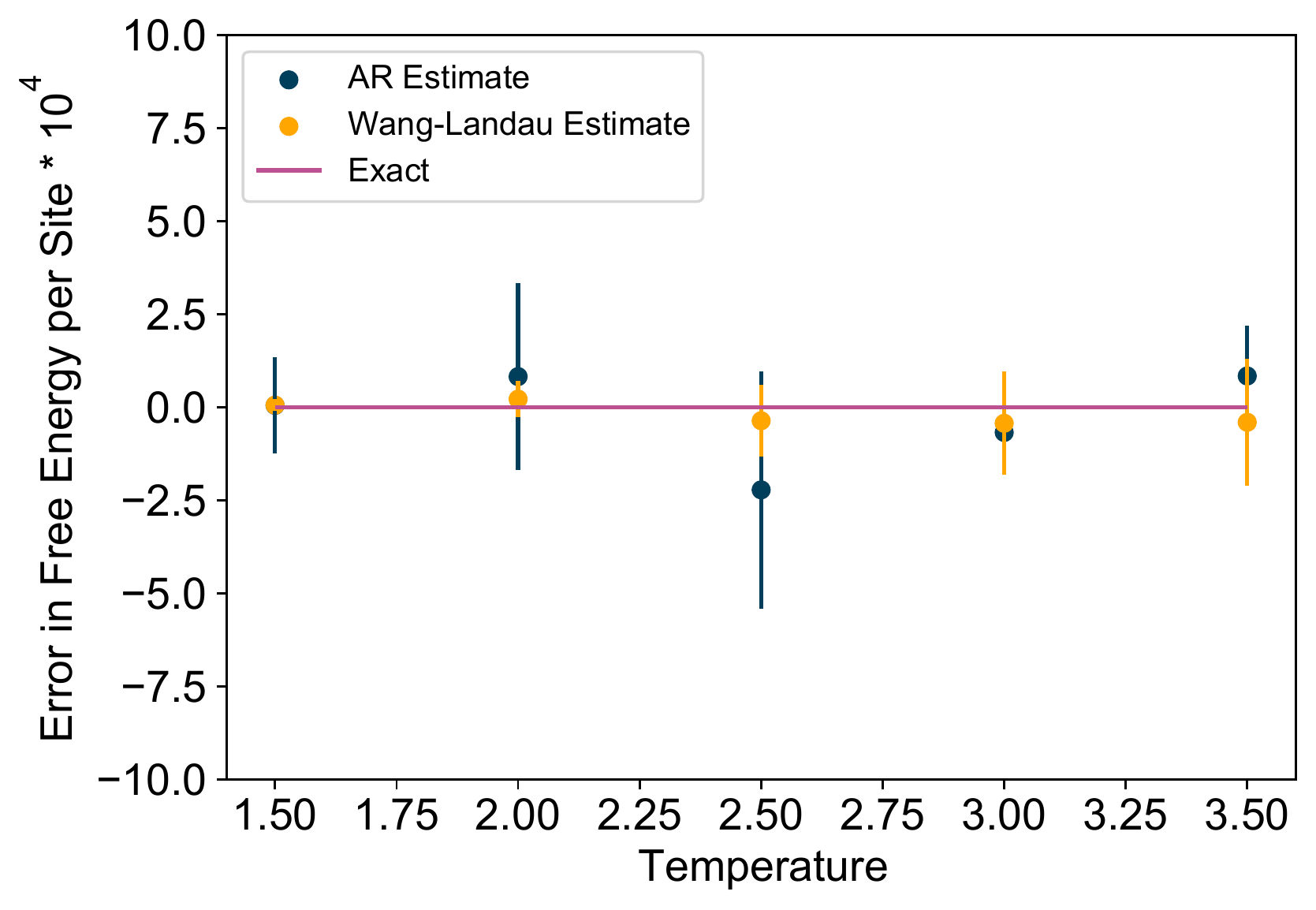} 
        \caption{} 
    \end{subfigure}

    \hfill
    \begin{subfigure}[t]{0.45\textwidth}
        \centering
        \includegraphics[width=\linewidth]{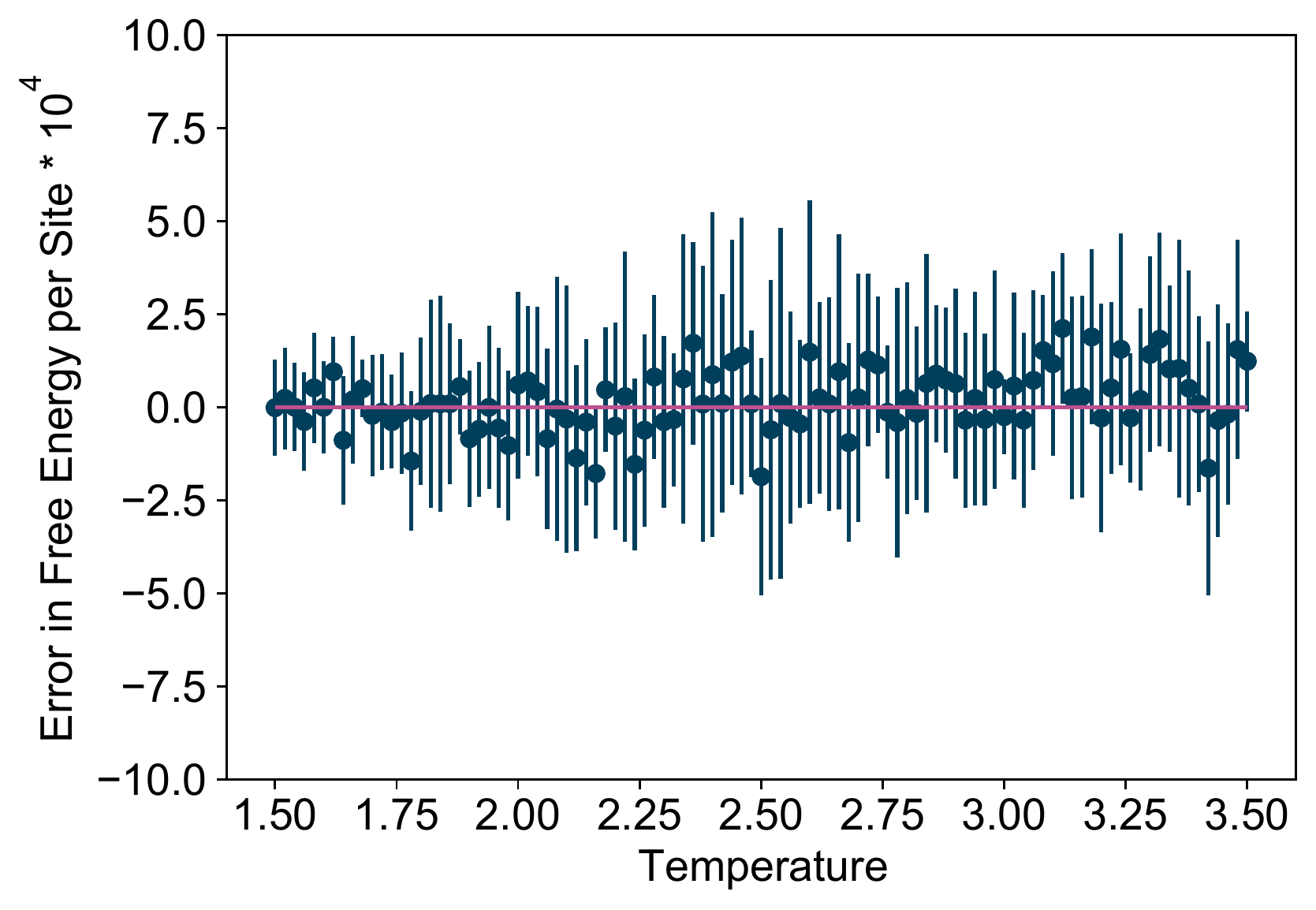} 
        \caption{} 
    \end{subfigure}
    
    \begin{subfigure}[t]{0.45\textwidth}
        \centering
        \includegraphics[width=\linewidth]{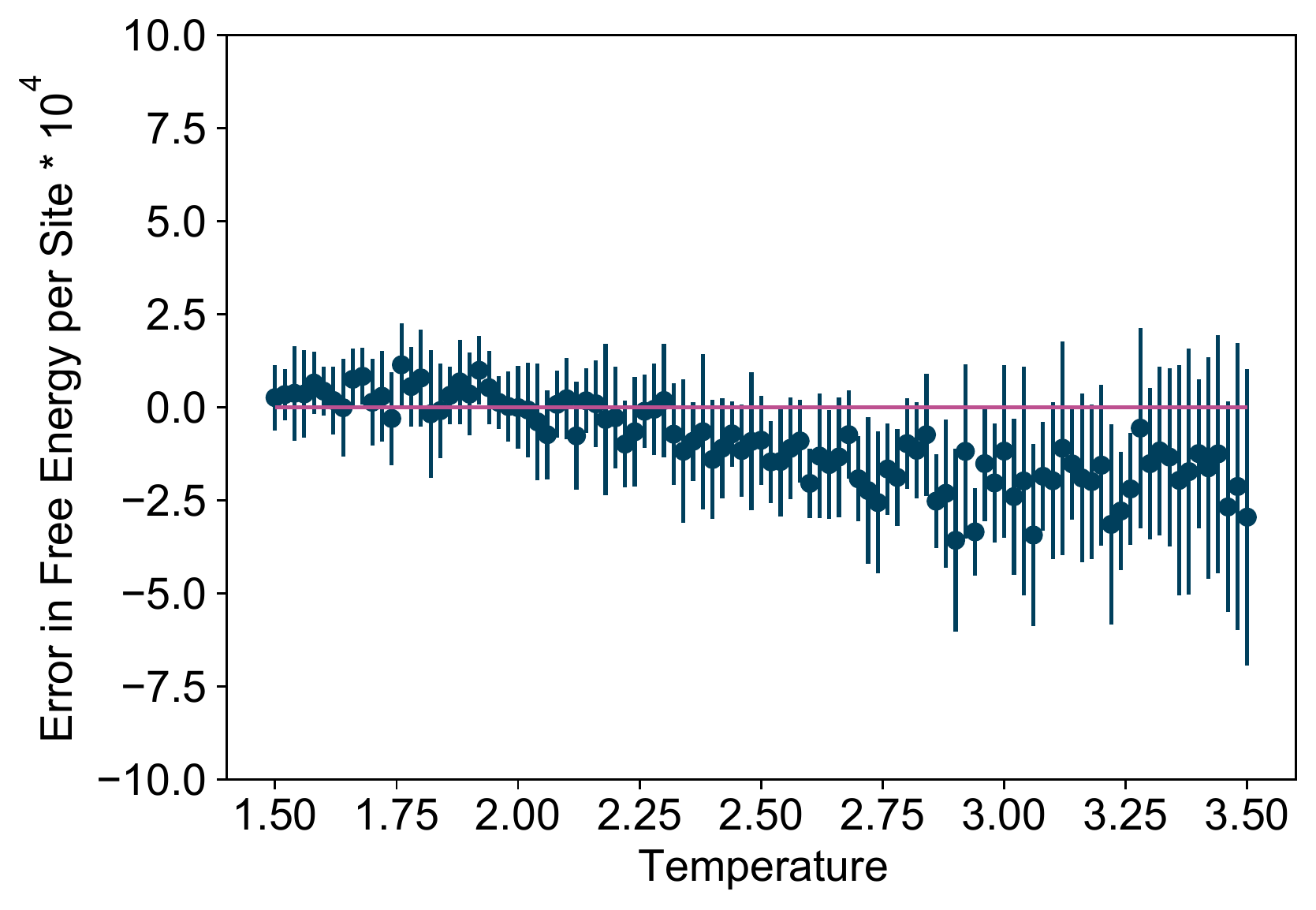} 
        \caption{} 
    \end{subfigure}
    
    \begin{subfigure}[t]{0.45\textwidth}
        \centering
        \includegraphics[width=\linewidth]{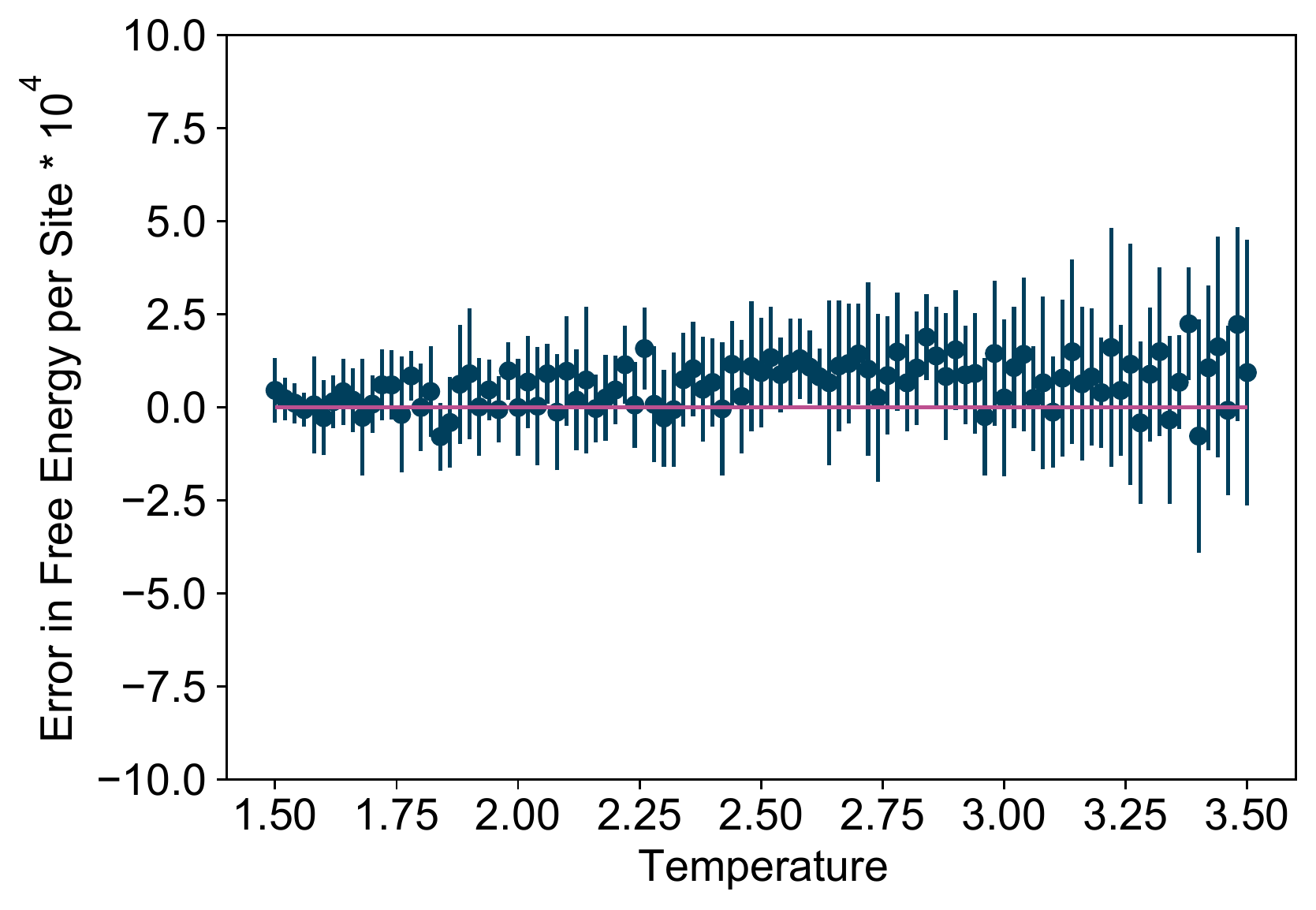} 
        \caption{} 
    \end{subfigure}
    
    \end{multicols}
    
    \caption{Errors in free energy per site computed by SEGAL through self-normalized importance sampling (SNIS). For each set of constraints, 10 independent free energy estimates are made using 2000 samples each. The mean (points) and standard deviation (bars) of these estimates are shown. (a) SEGAL and Wang-Landau free energies compared with exact values for $B=0.0$ case. (b-d) SEGAL estimates compared with Wang-Landau values for (b) $B=0.0$ , (c) $B=0.2$, and (d) $B=0.4$ cases. The benchmark Wang-Landau approach was an implementation of an algorithm by Beladinelli et al. \citep{Belardinelli2007} run for $10^{10}$ energy evaluations for each magnetic field.}
    \label{fig:IsingErrors}
\end{figure}

We determined the accuracy of SEGAL's free energy estimates when compared to the exact values at $B=0$ by computing the mean of the absolute error over temperatures in (1.5, 2.0, 2.5, 3.0, 3.5). SEGAL's estimates were made using 2000 samples each. We also determined the accuracy of a benchmark Wang-Landau method, computed in the same manner, when it was run at $B=0$ for varying numbers of energy evaluations. Error calculations were repeated over 10 independent trials for each method.

\begin{figure}[H]
\centering
\includegraphics[width=0.5\linewidth]{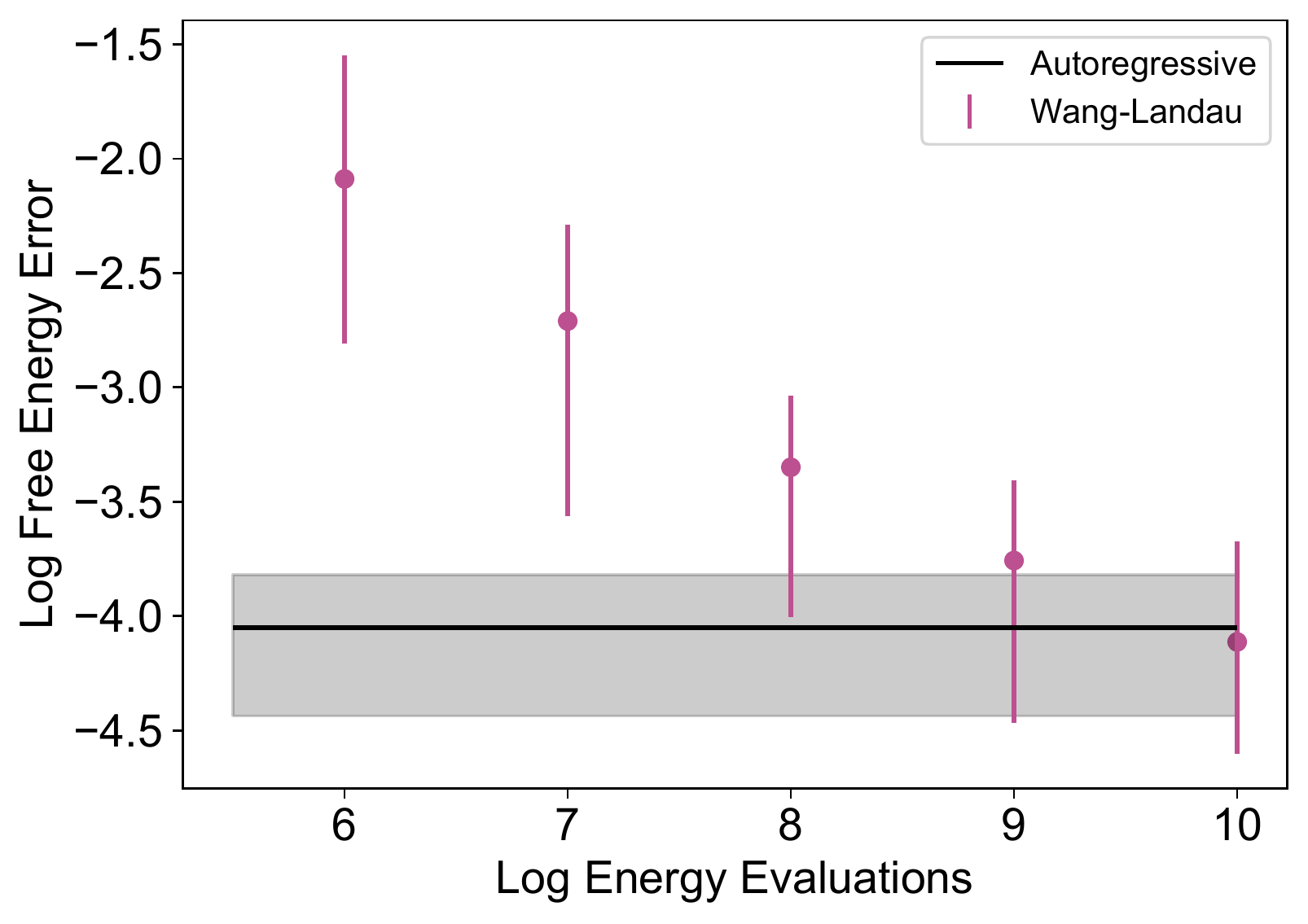}
\caption{Accuracy of SEGAL's estimated free energies compared with Wang-Landau benchmark at $B=0$. The energy evaluation cost of training SEGAL is $3\times10^{7}$. For each method, we report the mean error over 10 independent trials (points,horizontal line) as well as the max/min error range (error bars,shaded region) when compared with the exact values.}
\label{fig:IsingTiming}
\end{figure}

Usually, thermodynamic quantites obtained through derivatives such as the heat capacity and magnetic susceptibility are computed with Monte Carlo methods using the fluctuations of the samples. 

\begin{equation}
    C_{B} = [\langle E^{2}\rangle-\langle E \rangle^{2}]/k_{b}T
\end{equation}

\begin{equation}
    \chi = [\langle M^{2} \rangle-\langle M\rangle^{2}]/k_{b}T
\end{equation}

However, the differentiable structure of the function $P(\vec{S})$ allows for another strategy to obtain these derivatives. Treating $\chi$ as the example:

\begin{equation}
    \chi*k_{b}T = \frac{d}{dB}M(B,T) = \frac{d}{dB}\mathbf{E}_{AR}[M(\vec{S})]
\end{equation}

Using the same log-derivative trick found the methods section:

\begin{equation}
    \frac{d}{dB}\mathbf{E}_{AR}[M(\vec{S})] = \mathbf{E}_{AR}[M(\vec{S})\frac{d}{dB}log(P_{AR}(\vec{S}))]
\end{equation}

where $\frac{d}{dB}log(P_{AR}(\vec{S}))$ can be computed with automatic differentiation implemented in Pytorch. We report thermodynamic quantites obtained using this method below. We note that when heat capacities and susceptibilities are computed using automatic differentiation, they exhibit similar general trends to the fluctuation approach, but estimates are noisier and reduced in magnitude.

\begin{figure}[H]
    \centering
    \begin{multicols}{2}

    \hfill
    \begin{subfigure}[t]{0.45\textwidth}
        \centering
        \includegraphics[width=\linewidth]{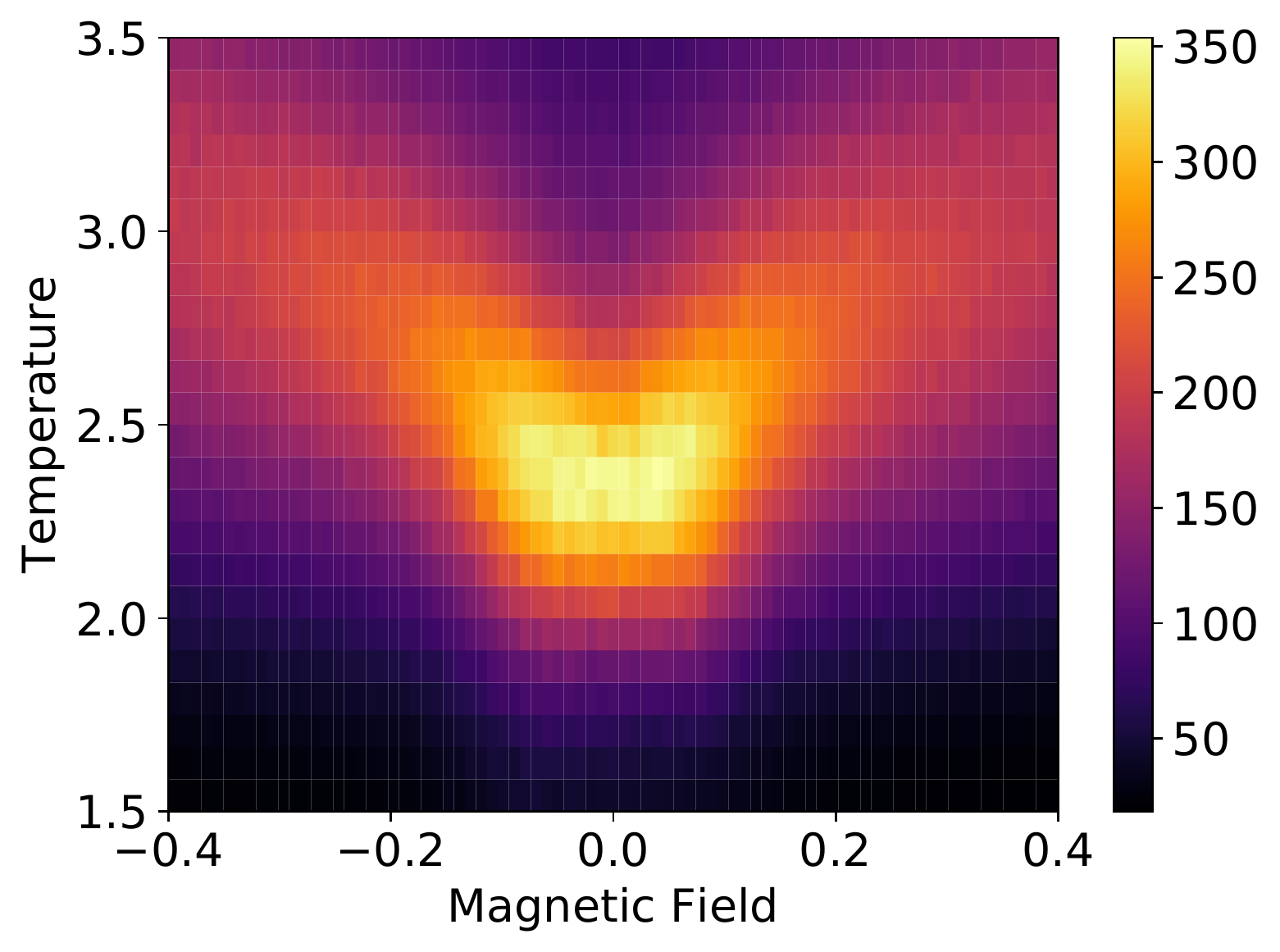} 
        \caption{} \label{fig:Ising_alloy_error_all}
    \end{subfigure}

    \hfill
    \begin{subfigure}[t]{0.45\textwidth}
        \centering
        \includegraphics[width=\linewidth]{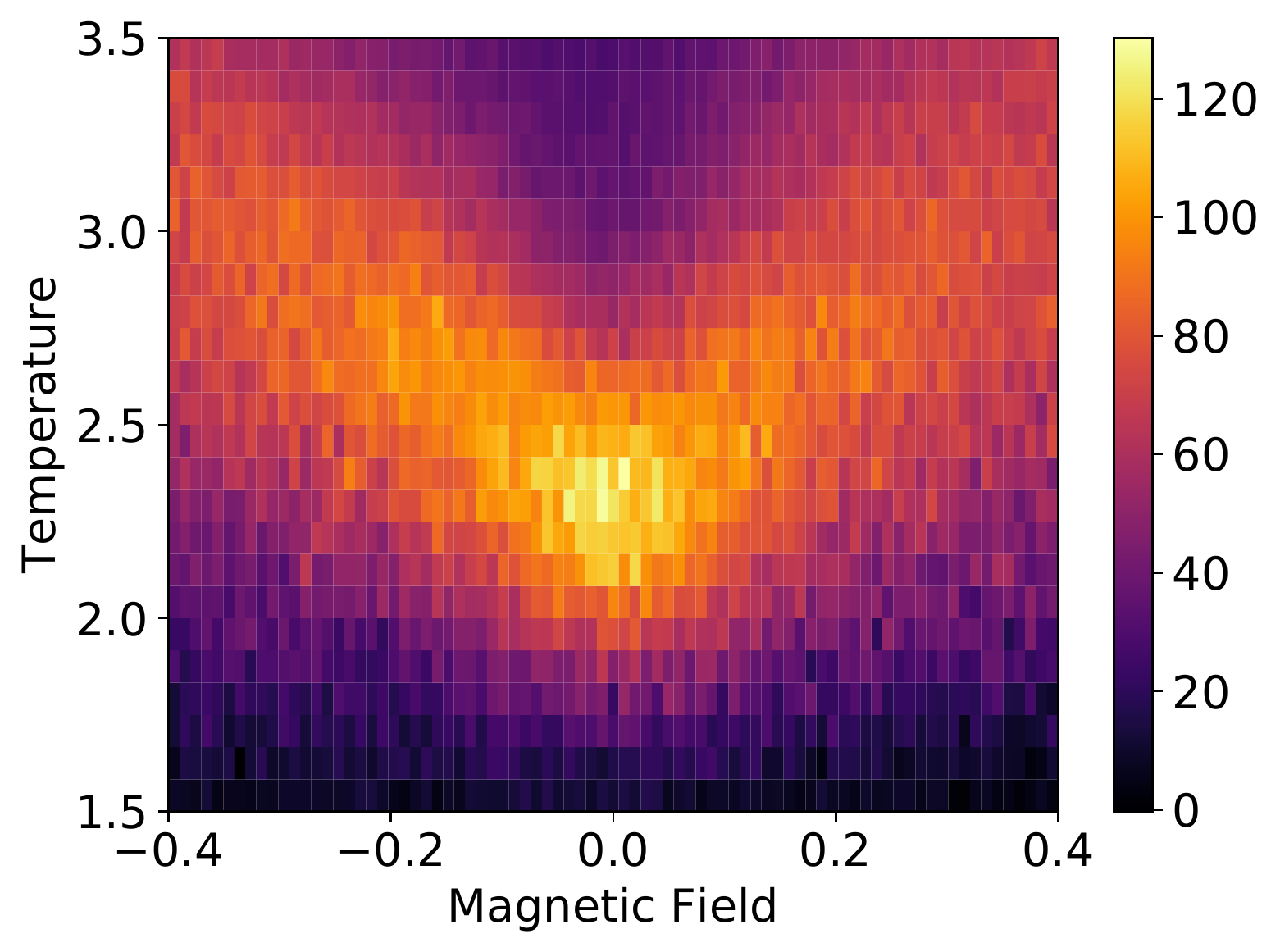} 
        \caption{} \label{fig:Ising_alloy_error_00}
    \end{subfigure}
    
    \begin{subfigure}[t]{0.45\textwidth}
        \centering
        \includegraphics[width=\linewidth]{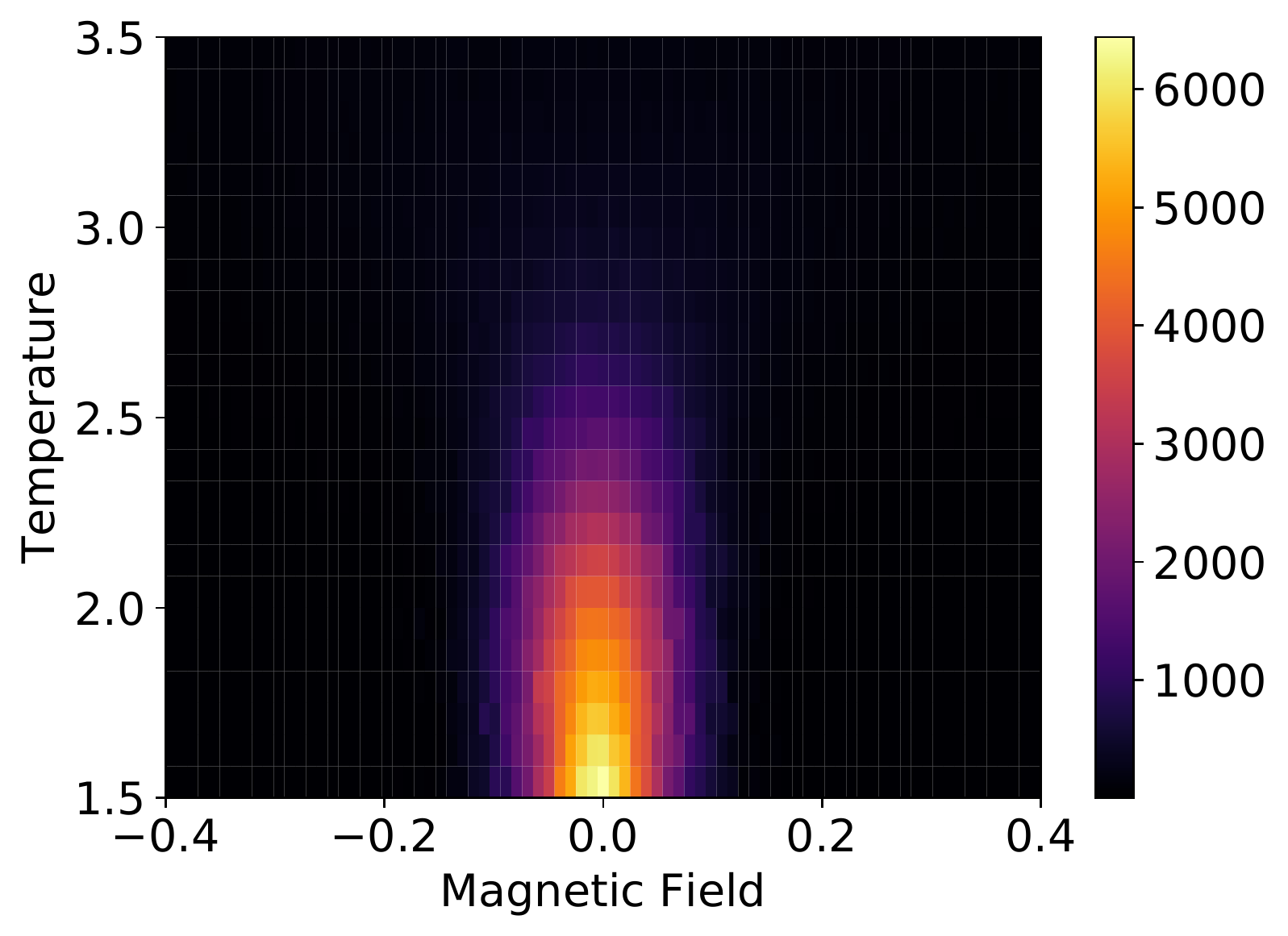} 
        \caption{} \label{fig:Ising_alloy_error_02}
    \end{subfigure}
    
    \begin{subfigure}[t]{0.45\textwidth}
        \centering
        \includegraphics[width=\linewidth]{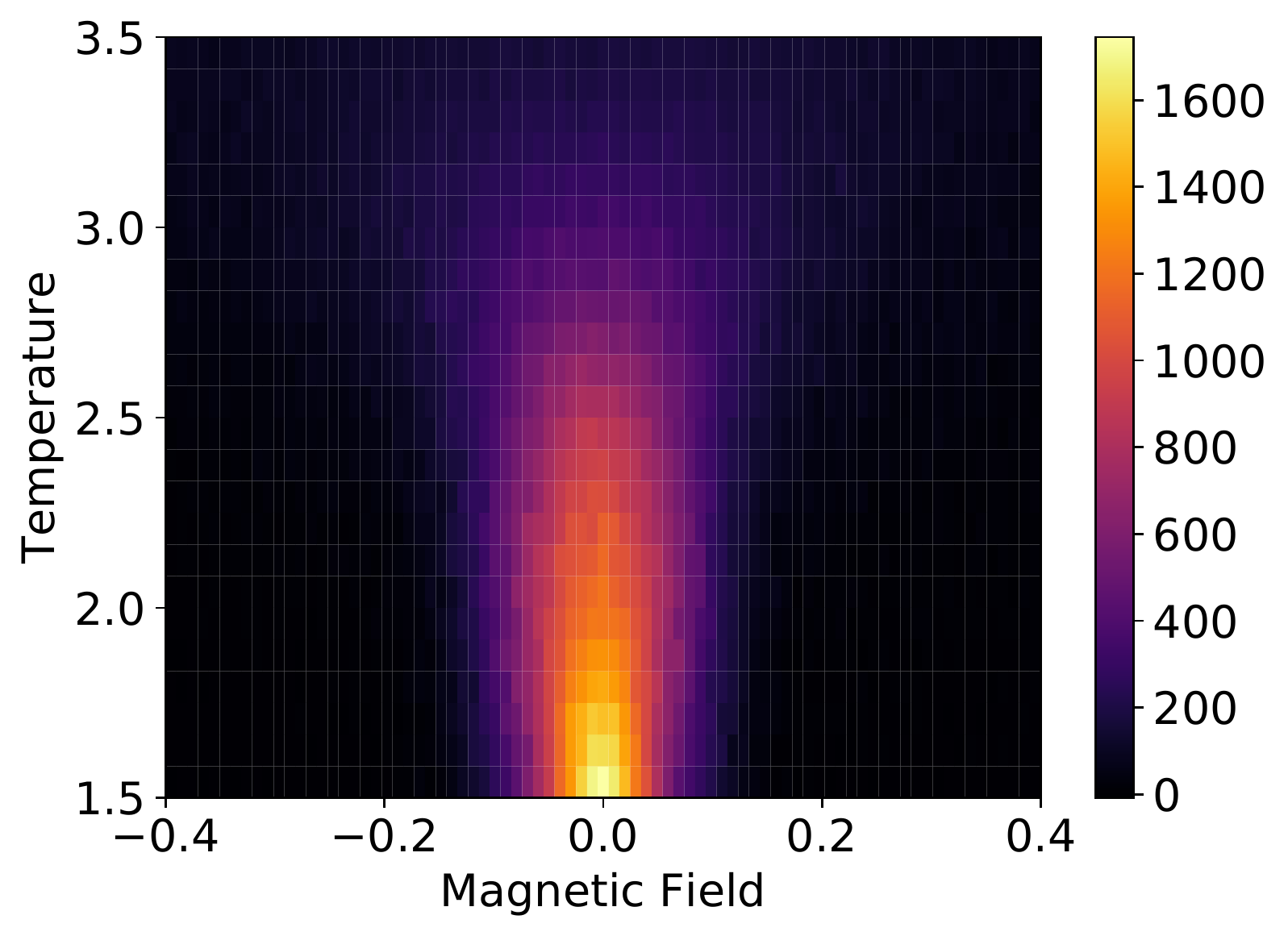} 
        \caption{} \label{fig:Ising_alloy_error_04}
    \end{subfigure}
    
    \end{multicols}
    
    \caption{Thermodynamic quantities computed using fluctuations of observables and neural network differentiation. All calculations use 10,000 samples at each set of constraints. (a) Heat capacity computed using fluctuations. (b) Heat capacity computed using neural network differentiation. (c) Magnetic susceptibility computed using fluctuations. (d) Magnetic susceptibility computed using neural network differentiation.}
    \label{fig:IsingAutoDiff}
\end{figure}

The Ising alloy models were composed of three layers as described above with tanh activation functions after the first two layers. Networks were trained for 22762 epochs with the rmsprop optimizer and a learning rate of $10^{-2.92}$. Each iteration contained 1250 total samples that were divided among 25 sets of conditions (50 samples per condition). The 25 conditions were chosen from all possible combinations of fields $B \in [-0.4,-0.2,0.0,0.2,0.4] $ and 5 randomly chosen temperatures in range $1.5 < T_{i} < 3.5$. Hyperparameters were optimized using SigOpt \citep{SigOpt2019} and all neural networks implementations were build with PyTorch \citep{PyTorch2019}. 

\subsection{CuAu Alloy Experimental Details}
\label{coppergold}

\begin{figure}[H]
    \centering
    \begin{subfigure}[t]{0.45\textwidth}
        \centering
        \includegraphics[width=\linewidth]{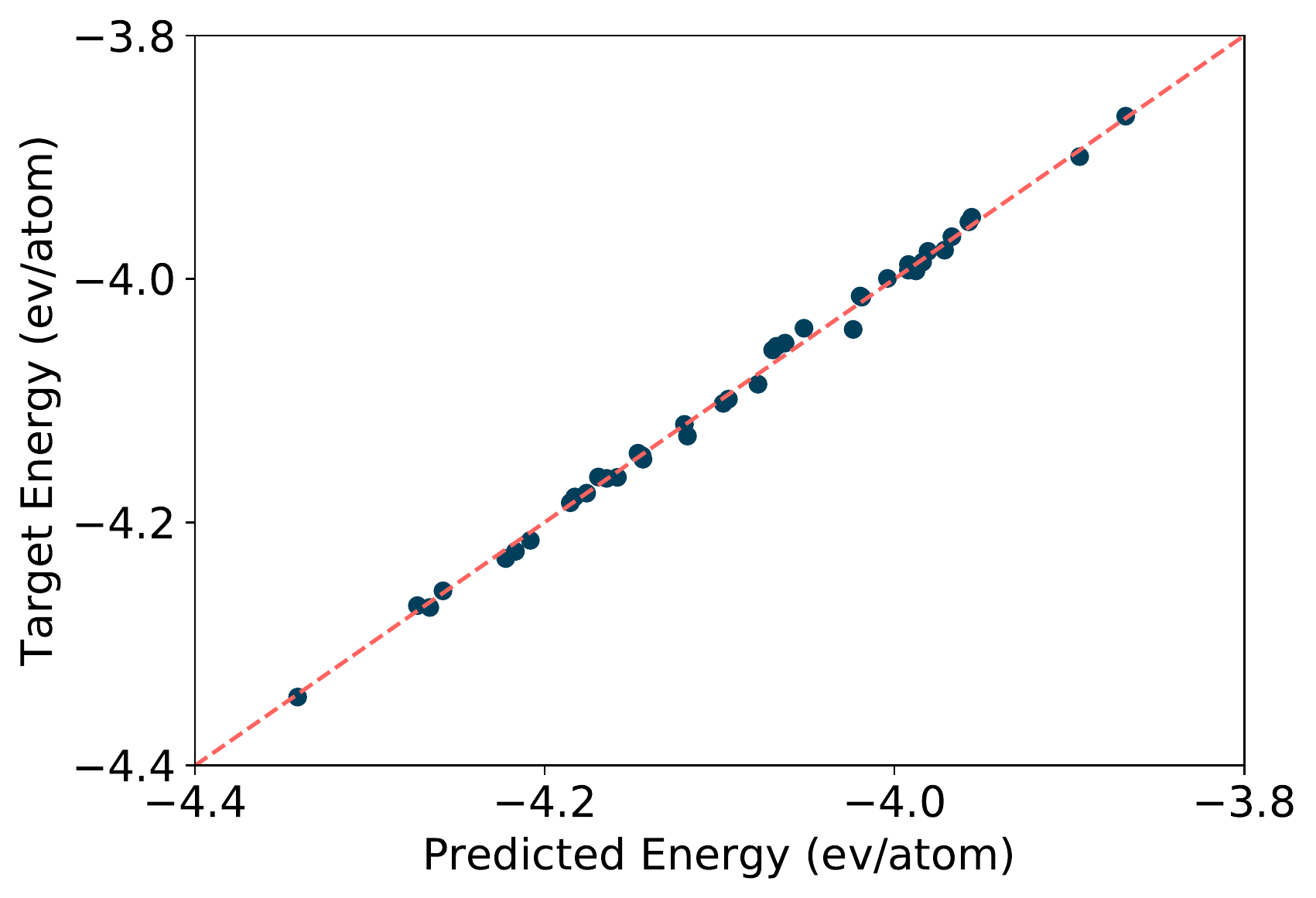} 
        \caption{} \label{fig:CuAu_energy_regression}
    \end{subfigure}
    \caption{Fit of CuAu cluster expansion to total energy.}
    \label{fig:CuAu_ClusterExpansion}
\end{figure}

The CuAu cluster expansion was obtained using CLEASE \citep{Chang2019} on an fcc structure with cell parameter 3.8 {\AA}. Max cluster diameters for two, three, and four body clusters were 6.0 {\AA}, 4.5 {\AA}, and 4.5 {\AA} respectively. The expansion was fit with LOOCV and L2 regularization. 

The autoregressive models were composed of three layers with sigmoid activation. Networks were trained for 5000 epochs with the Adam optimizer and a learning rate of $10^{-3}$. Each iteration contained 200 total samples that were divided among 4 sets of conditions (50 samples per condition). The 4 conditions were chosen using random chemical potentials $\Delta \mu \in [-0.24,0.24] $ and an annealed temperature schedule $T=3000*(0.999)^{iter}$.

\begin{figure}[H]
\centering
\includegraphics[width=0.5\linewidth]{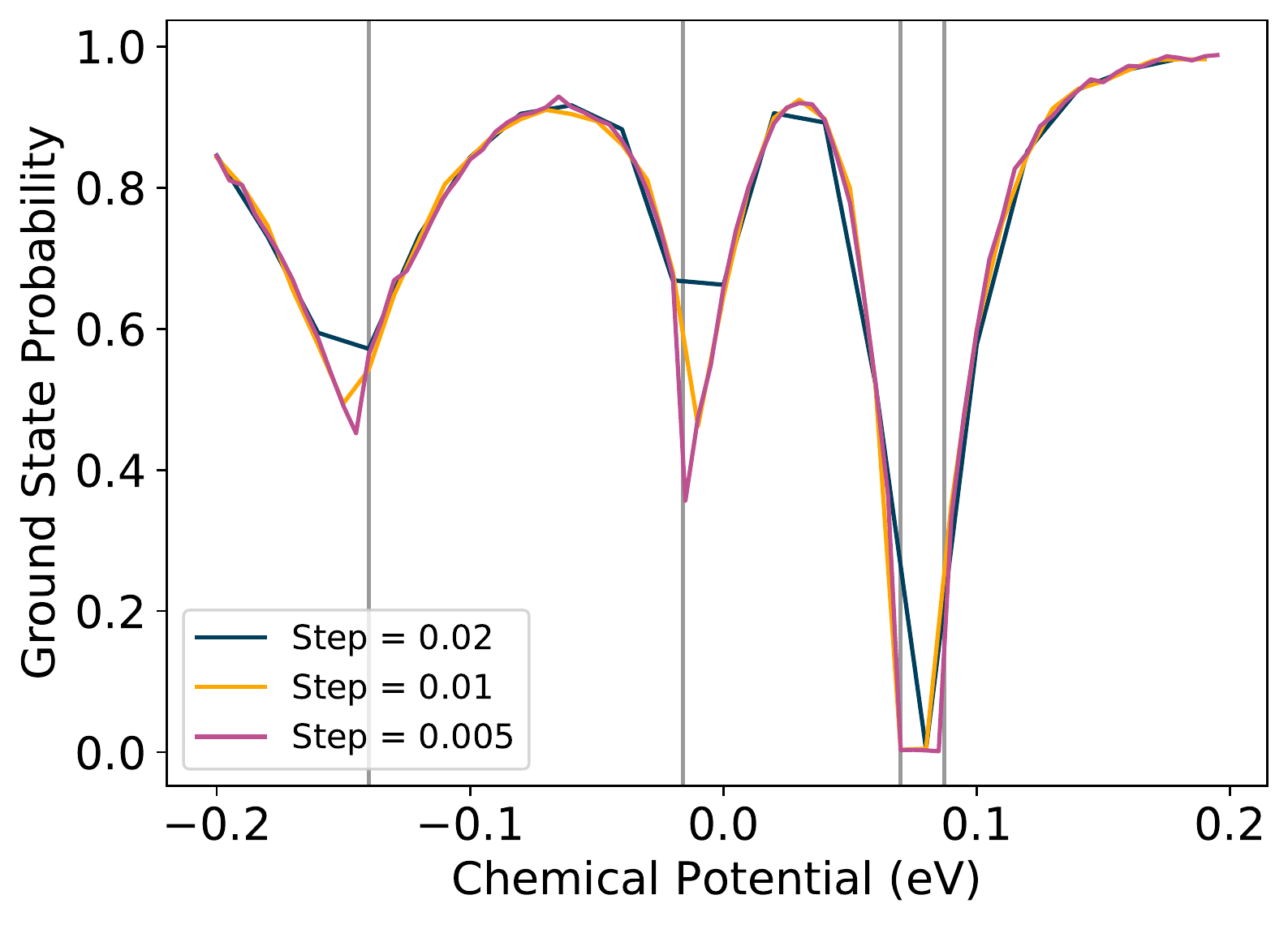}
\caption{Probability of sampling the grand potential minima as a function of chemical potential. Each $\Delta\mu$ is evaluated using a batch of 2500 samples. Varying step sizes are plotted to show the effects of approaching the phase transitions with increasing resolution. The minium probability over all resolutions is 3/2500 and occurs at a $\Delta\mu$ value of 0.085. }
\label{fig:CuAu_GSProb}
\end{figure}

\subsection{AgPd Alloy Experimental Details}
\label{silverpalladium}

CE:
The AgPd cluster expansion was obtained using CLEASE \citep{Chang2019} on an fcc structure with cell parameter 4.09 {\AA}. Max cluster diameters for two, three, and four body clusters were 8.0 {\AA}, 6.5 {\AA}, and 5.5 {\AA} respectively. Of the 625 structures in the ICET example, 613 were correctly added to the CLEASE database. The expansion was fit with k-fold cross validation and L2 regularization. 

CGC:
Crysal Graph Convolutions were trained with a (80,10,10) (training, validation, test) split. Training continued for 300 epochs with a batch size of 100. The objective was optimized with Adam. The learning rate was initially set to 0.01 and reduced to 0.001 after 100 epochs. Various parameters of the convolutional structure are listed below.

\begin{center}
\begin{tabular}{| c | c |} 
\hline
hidden atom features in conv layers & 64 \\ 
\hline
hidden features after pooling & 64\\ 
\hline
number of conv layers & 3\\ 
\hline
number of hidden layers after pooling & 2\\ 
\hline
\end{tabular}
\end{center}

Experimental Details:
 Both 27-site AgPd models were composed of three layers as described above with sigmoid activation functions after the first two layers. Networks were trained for 17229 epochs with the Adam optimizer and a learning rate of $10^{-2.52}$. Each iteration contained 1250 total samples that were divided among 25 sets of conditions (50 samples per condition). The 25 conditions were chosen from all possible combinations of chemical potentials $ \Delta \mu \in [-0.4,-0.2,0.0,0.2,0.4] $ and 5 randomly chosen temperatures in range $200 K < T_{i} < 900 K$. Hyperparameters were optimized using SigOpt \citep{SigOpt2019} and all neural networks implementations were build with PyTorch \citep{PyTorch2019}. Benchmark SGC MCMC simulations were run for 1000 sweeps (27 MC steps each) with one sample recorded for each sweep. At each temperature, we performed importance sampling from SEGAL and ran a MCMC simulation for 41 values of $\Delta\mu$ ranging from -0.4 to 0.4.

\begin{figure}[H]
    \centering
    \begin{subfigure}[t]{0.45\textwidth}
        \centering
        \includegraphics[width=\linewidth]{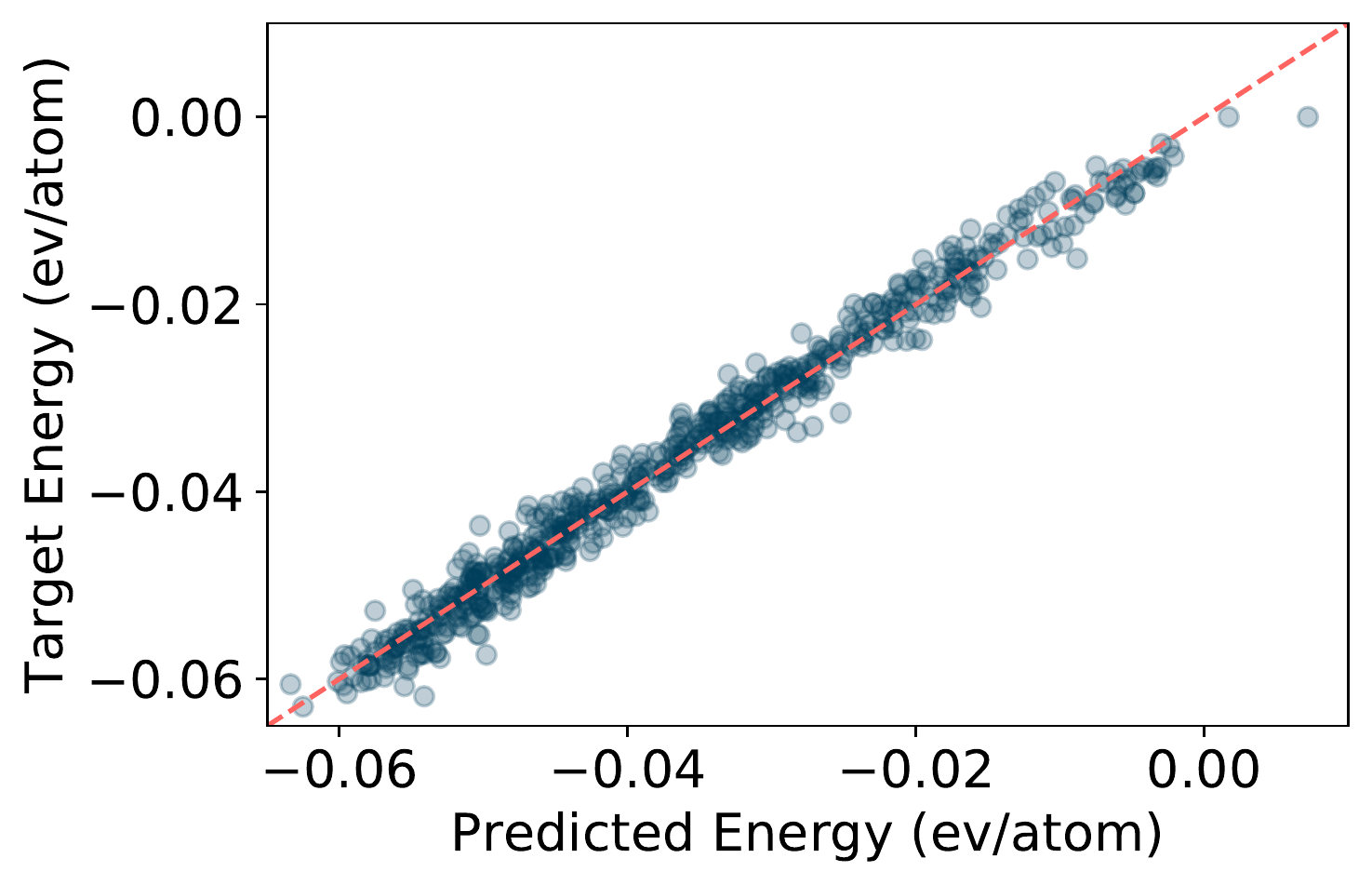} 
        \caption{} \label{fig:AgPd_ClusterExpansion}
    \end{subfigure}
    \hfill
    \begin{subfigure}[t]{0.45\textwidth}
        \centering
        \includegraphics[width=\linewidth]{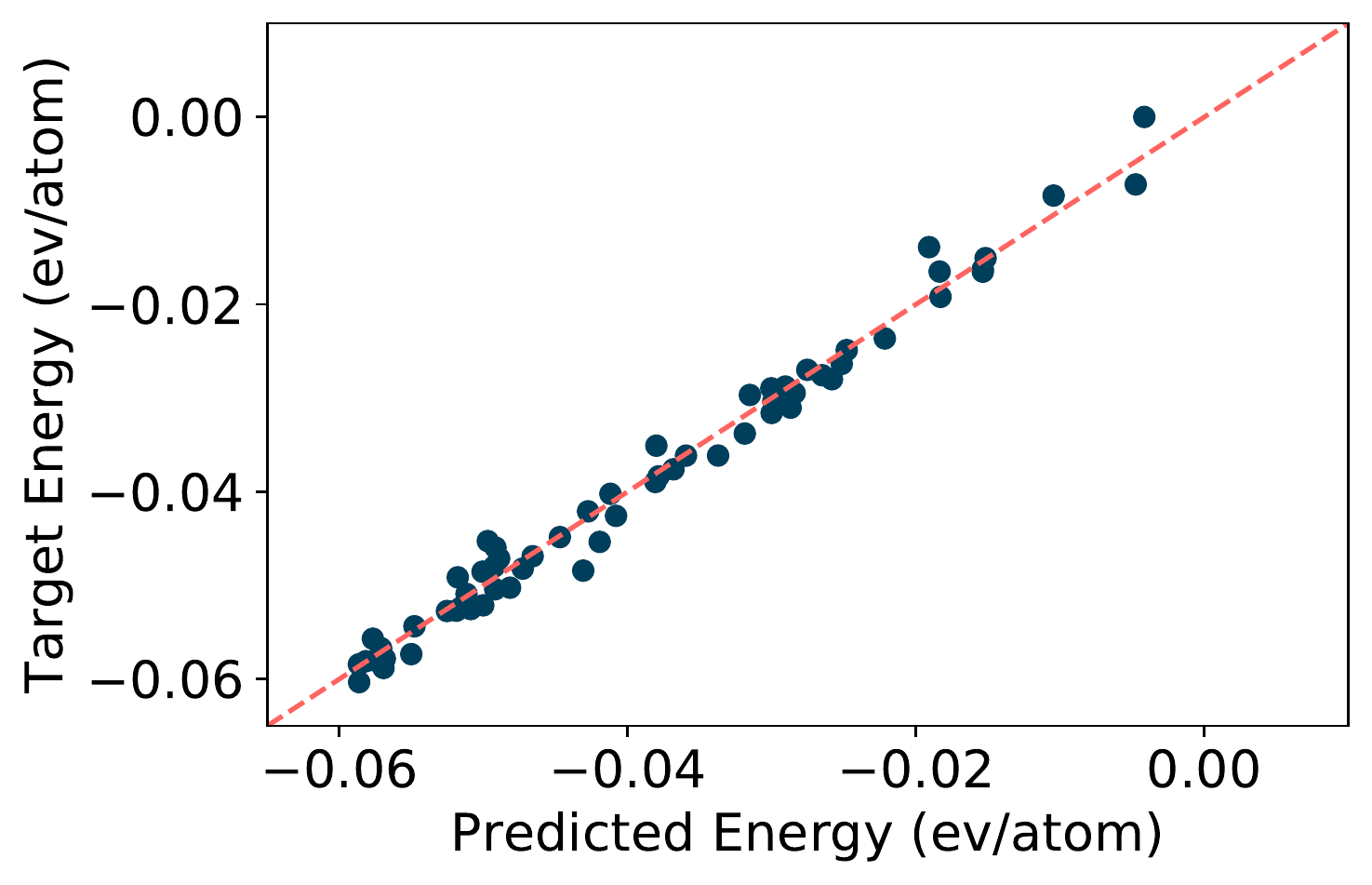}
        \caption{} \label{fig:AgPd_CGC}
    \end{subfigure}
    \caption{AgPd models for $U$. (a) Fit of AgPd cluster expansion to formation energy. (b) Performance of Crystal Graph Convolutional fit of formation energy on test set.}
    \label{fig:AgPd Energy Functions}
\end{figure}

\begin{figure}[H]
    \centering
    \begin{multicols}{2}

    \hfill
    \begin{subfigure}[t]{0.45\textwidth}
        \centering
        \includegraphics[width=\linewidth]{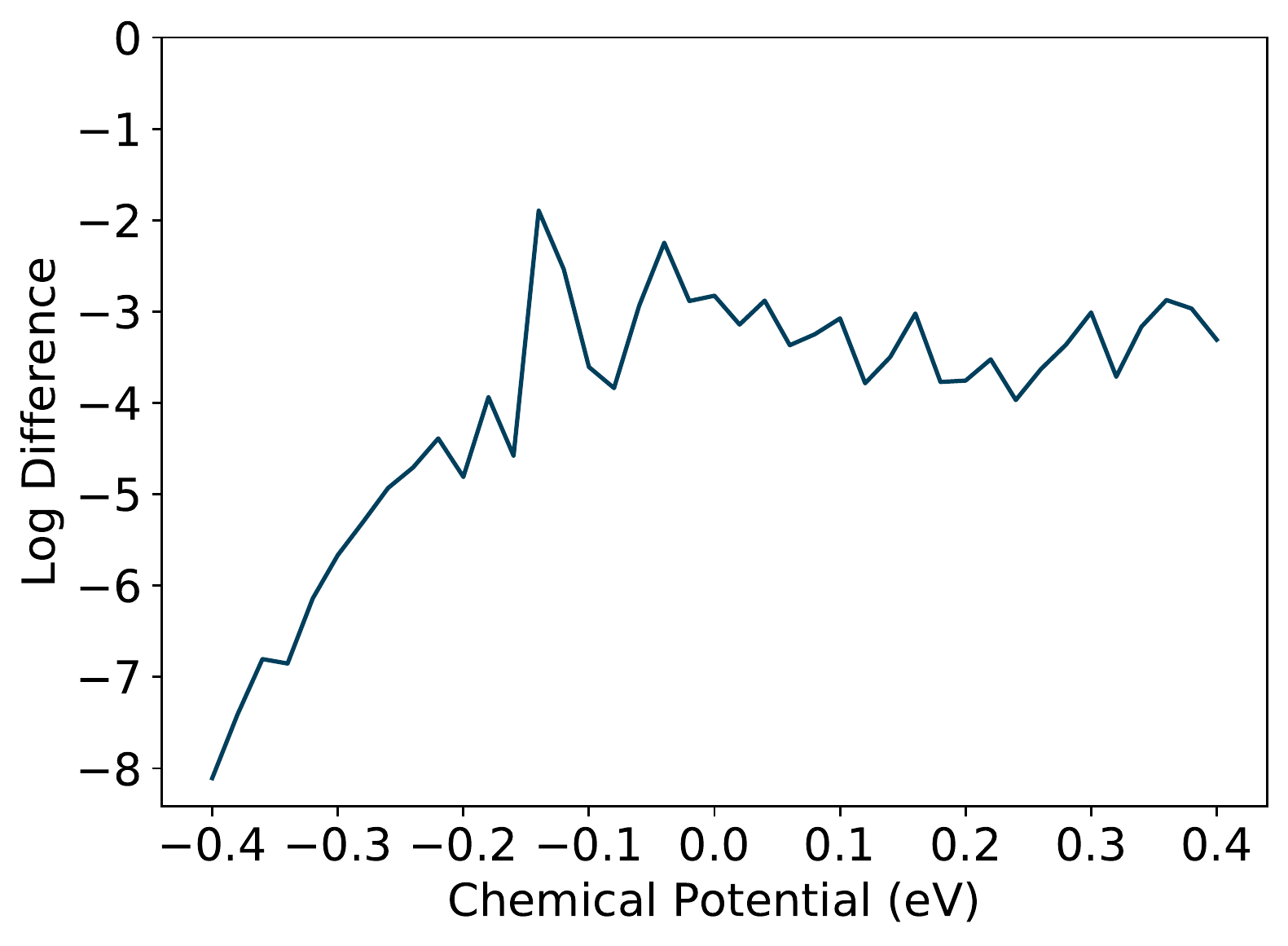} 
        \caption{} \label{fig:AgPd_error_1}
    \end{subfigure}

    \hfill
    \begin{subfigure}[t]{0.45\textwidth}
        \centering
        \includegraphics[width=\linewidth]{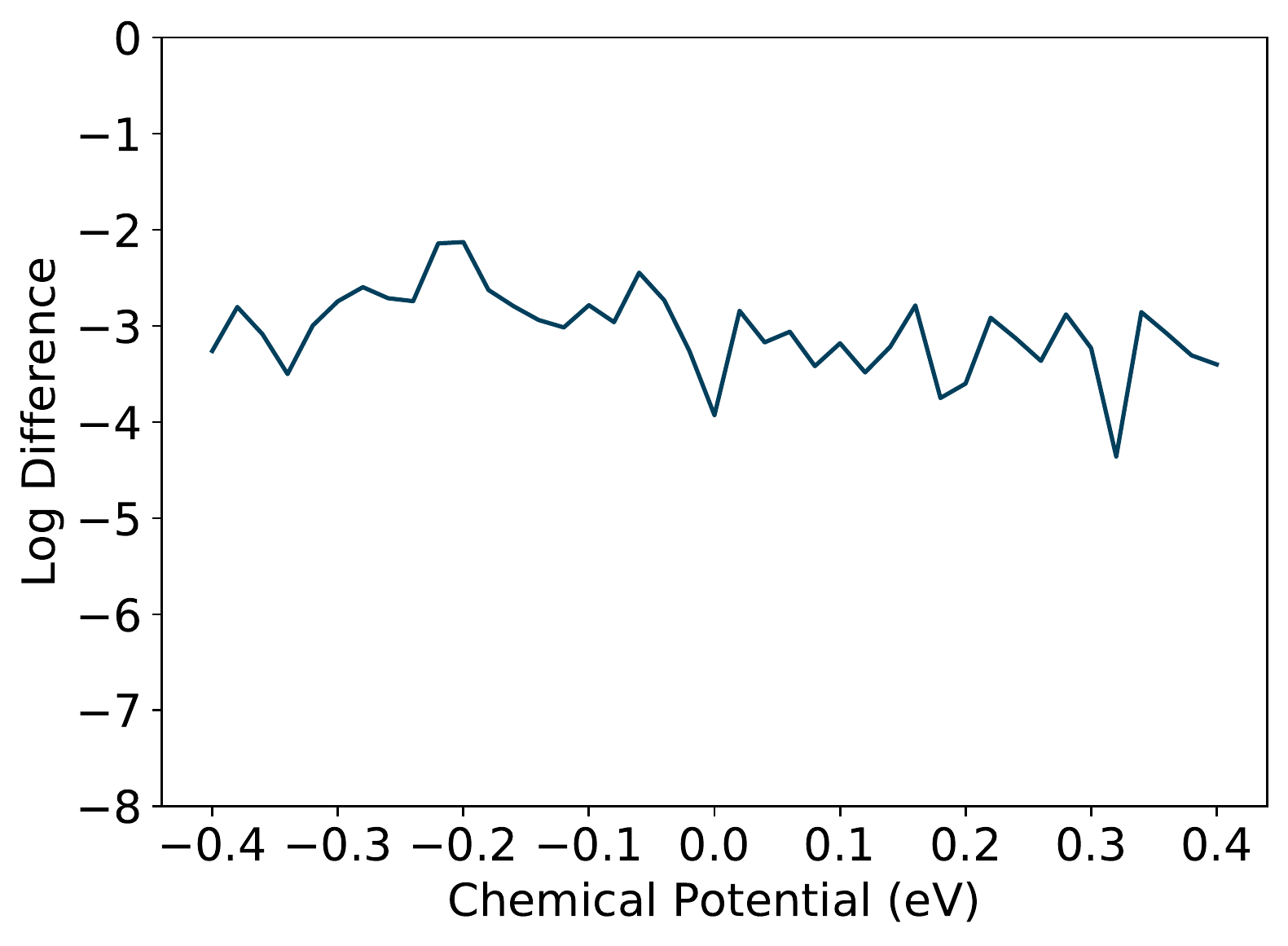} 
        \caption{} \label{fig:AgPd_error_2}
    \end{subfigure}
    
    \begin{subfigure}[t]{0.45\textwidth}
        \centering
        \includegraphics[width=\linewidth]{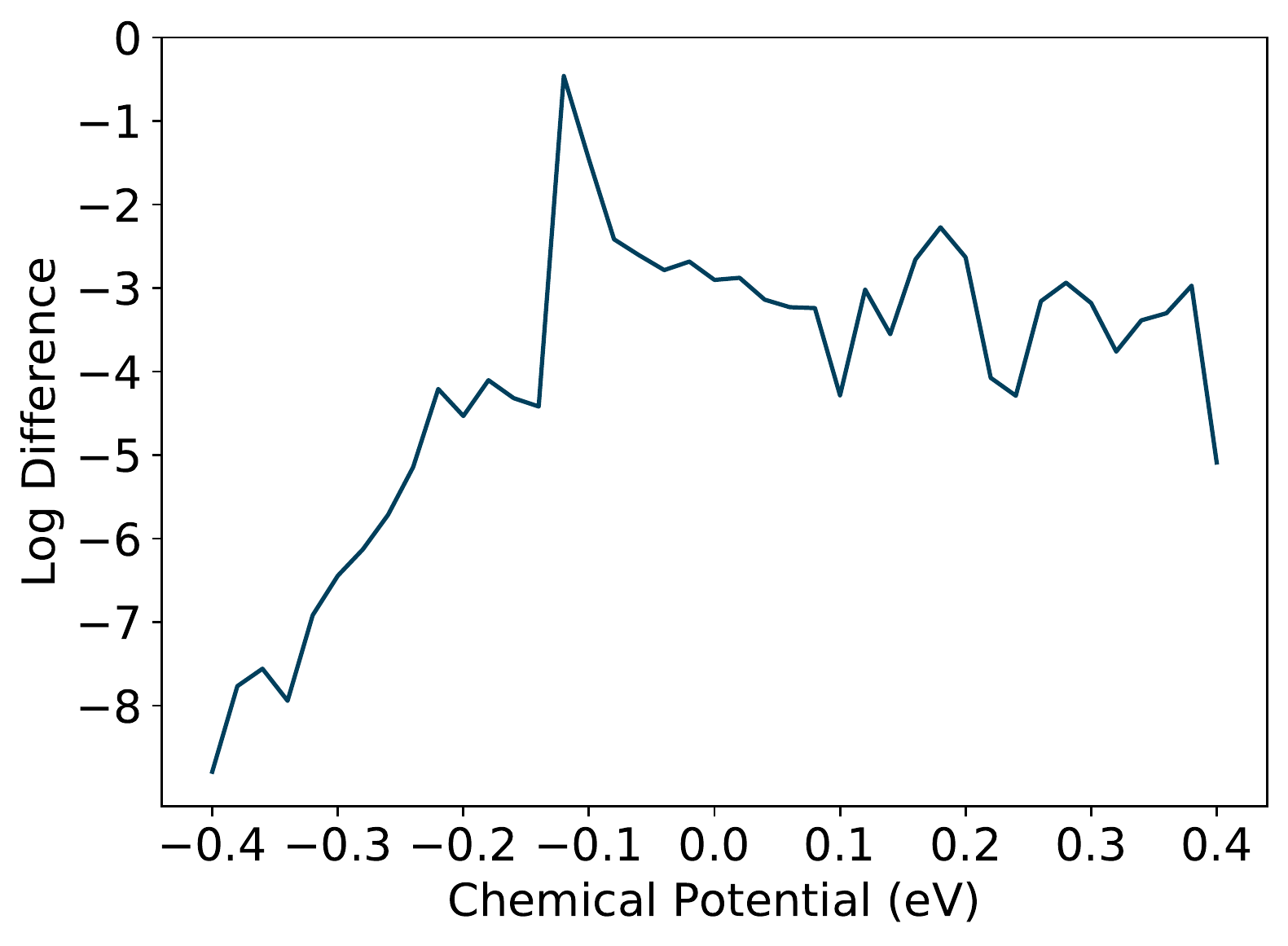} 
        \caption{} \label{fig:AgPd_error_3}
    \end{subfigure}
    
    \begin{subfigure}[t]{0.45\textwidth}
        \centering
        \includegraphics[width=\linewidth]{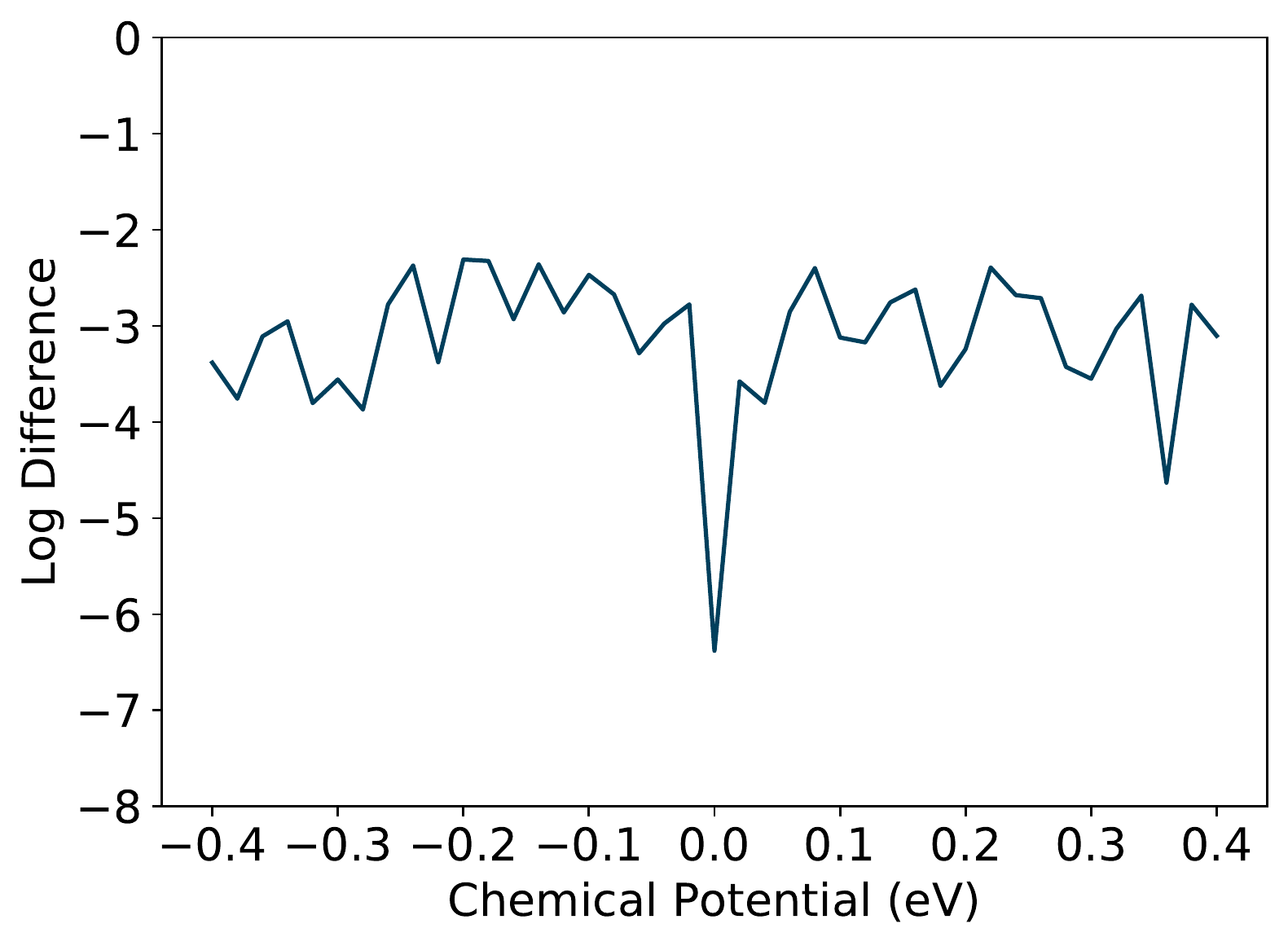} 
        \caption{} \label{fig:AgPd_error_4}
    \end{subfigure}
    
    \end{multicols}
    
    \caption{Deviations in composition between SEGAL estimates and SGC MCMC simulations. (a) Cluster Expansion at 250 K. (b) Cluster Expansion at 750 K. (c) Crystal Graph Convolution at 250 K. (d) Crystal Graph Convolution at 750 K. Below the peak of miscibility gap, deviations peak at the discontinuity in composition due to SEGAL's performance near the phase transition.}
    \label{fig:AgPd Errors}
\end{figure}

\subsection{Phase Diagrams Experimental Details}
\label{phasediagrams}

\begin{figure}[H]
    \centering
    \begin{subfigure}[t]{0.45\textwidth}
        \centering
        \includegraphics[width=\linewidth]{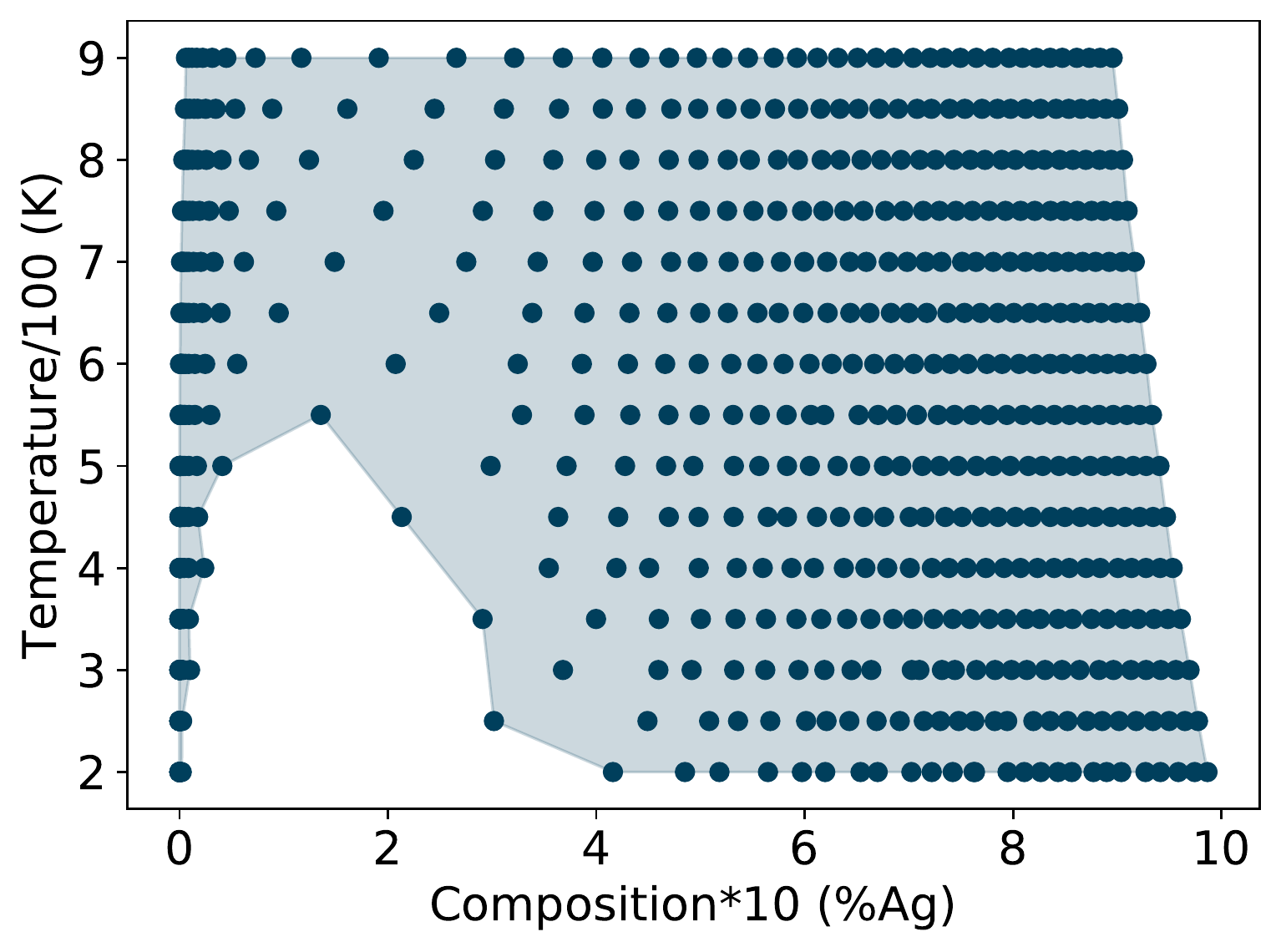} 
        \caption{} \label{fig:AgPd_AlphaShape}
    \end{subfigure}
    \hfill
    \begin{subfigure}[t]{0.45\textwidth}
        \centering
        \includegraphics[width=\linewidth]{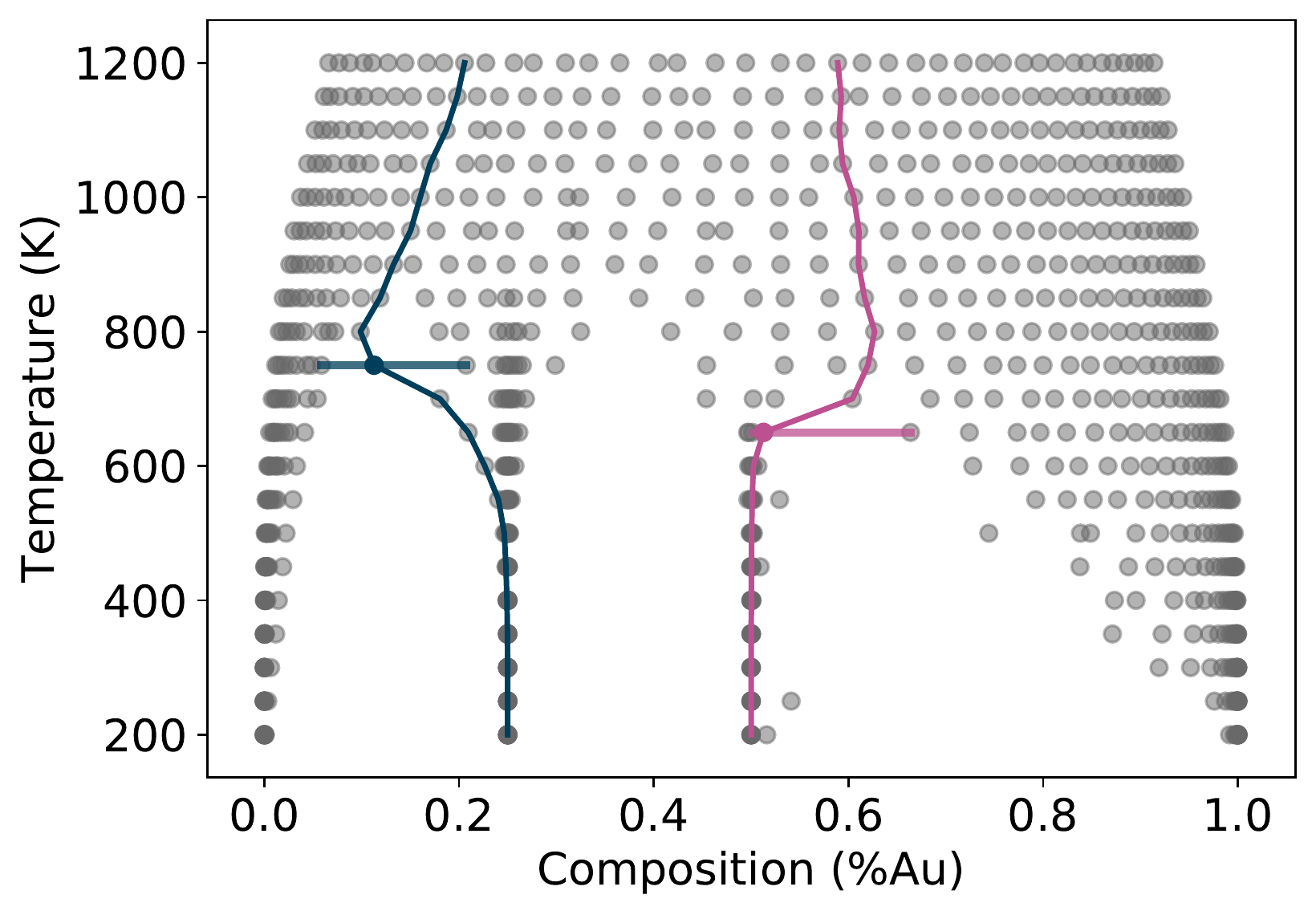}
        \caption{} \label{fig:CuAu_HeatCap}
    \end{subfigure}
    \caption{(a) Alpha shape for 125-site AgPd SEGAL model obtained with alpha value 1.2 \citep{Bellock2021}. Compositions and temperatures are set to similar scales to stabalize the alpha shape algorithm. Figure shows importance sampling estimates of composition for 41 values of $\Delta\mu \in$ [-0.4,0.4] and 15 temperatures $\in$ [200,900]. Each estimate is computed using 5,000 SEGAL samples. (b) Procedure for identifying order-disorder two-phase region for CuAu system. Vertical paths are at constant $\Delta\mu$ (blue: $\Delta\mu$ = -0.132, purple: $\Delta\mu$ = 0.024), and points denote the critical temperature $T_{C}$. Horizontal lines show the approximated bounds of the two-phase region. Figure shows importance sampling estimates of composition for 41 values of $\Delta\mu \in$ [-0.24,0.24] and 21 temperatures $\in$ [200,1200]. Each estimate is computed using 5,000 SEGAL samples.}
    \label{fig:Large NESS}
\end{figure}

\begin{figure}[H]
    \centering
    \begin{subfigure}[t]{0.45\textwidth}
        \centering
        \includegraphics[width=\linewidth]{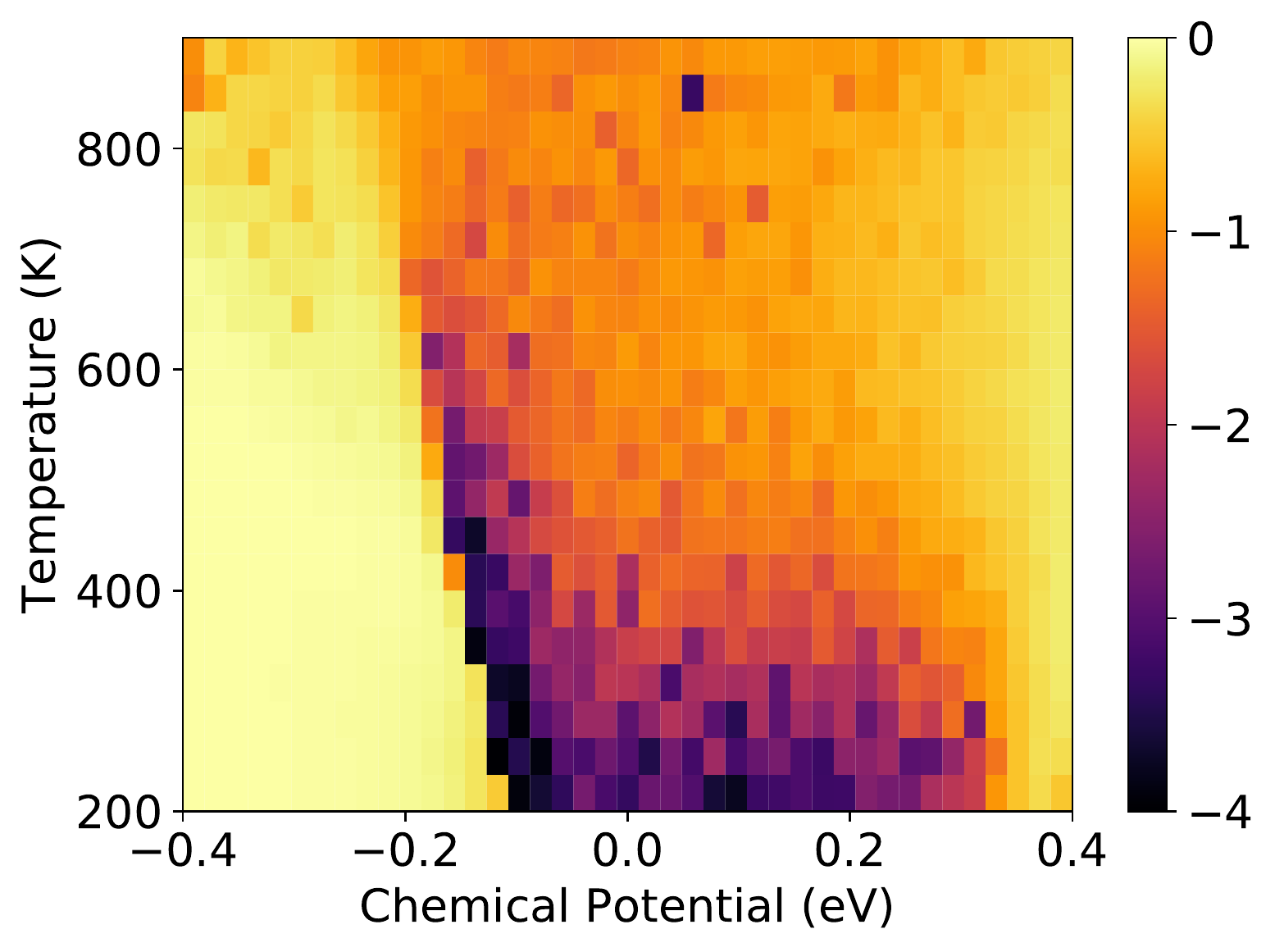} 
        \caption{} \label{fig:AgPd_Large_NESS}
    \end{subfigure}
    \hfill
    \begin{subfigure}[t]{0.45\textwidth}
        \centering
        \includegraphics[width=\linewidth]{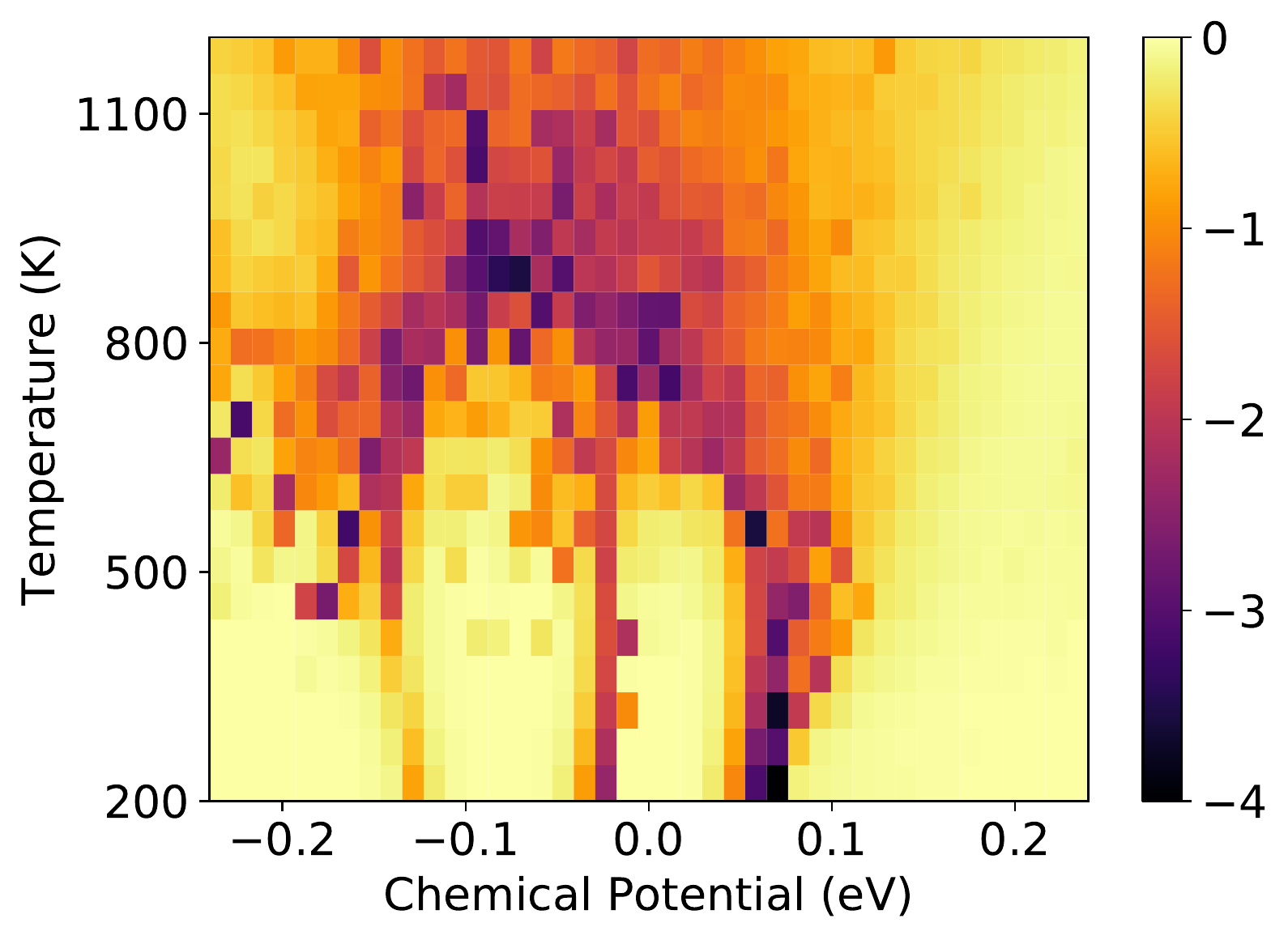}
        \caption{} \label{fig:CuAu_Large_NESS}
    \end{subfigure}
    \caption{Normalized Effective Sample Size of (a) 125-site AgPd model and (b) 128-site AuCu model reported in Log10 scale. All estimates are made using 10,000 samples from SEGAL.}
    \label{fig:LargeNESS}
\end{figure}

Experimental Details:
The 125-site AgPd model had 2 layers and tanh activation functions. Networks were trained for 13285 epochs with a learning rate of $10^{-2.68}$ and the adam optimizer. Each batch contained 500 total samples that were divided among 25 sets of conditions that were chosen in the same manner as the 27-site model. 

The 128-site CuAu SEGAL model was trained with the same settings as the 27-site AgPd model. The temperature training range was [200 K,1200 K], and the $\Delta\mu$ values were chosen as [-0.2,-0.08,0.0,0.08,0.2]. 

Metadynamics simulations are run using the CLEASE \citep{Chang2019} package. For AgPd, simulations are run for 15 temperatures from 200 K to 900 K. The flatness criteria is set to 0.8. The modification factor is initialized at 1000, and the simulation continues until the modification factor is less than 0.0005, where the modification factor is reduced by a factor of 2.0 after the flatness criteria has been reached. For CuAu, simulations are run for 41 temperatures from 200 K to 1200 K. The flatness criteria is set to 0.8. The modification factor is initialized at 10000, and the simulation continues until the modification factor is less than 0.0001.
\end{document}